\documentclass[oldversion]{aa} % double column
\usepackage{graphicx}
\usepackage{longtable}
\usepackage{txfonts}
\usepackage{rotating}
\usepackage{lscape}
\usepackage{color}
\newcommand{\gsim}{\;\lower.6ex\hbox{$\sim$}\kern-7.75pt\raise.65ex\hbox{$>$}\;}
\newcommand{\lsim}{\;\lower.6ex\hbox{$\sim$}\kern-7.75pt\raise.65ex\hbox{$<$}\;}

\begin{document}
\title{The normal chemistry of multiple stellar populations in the dense globular 
cluster NGC~6093 (M~80)\thanks{Based on observations collected at 
ESO telescopes under programme 083.D-0208}\fnmsep\thanks{
   Tables 2, 3, 5, 6, 7, 8 and 9 are only available in electronic form at the CDS via anonymous
   ftp to {\tt cdsarc.u-strasbg.fr} (130.79.128.5) or via
   {\tt http://cdsweb.u-strasbg.fr/cgi-bin/qcat?J/A+A/???/???}}
 }

\author{
E. Carretta\inst{1},
A. Bragaglia\inst{1},
R.G. Gratton\inst{2},
V. D'Orazi\inst{2,3,4},
S. Lucatello\inst{2},
A. Sollima\inst{1},
Y. Momany\inst{2,5},
G. Catanzaro\inst{6},
\and
F. Leone\inst{7}
}

\authorrunning{E. Carretta et al.}
\titlerunning{Multiple stellar populations in NGC~6093}

\offprints{E. Carretta, eugenio.carretta@oabo.inaf.it}

\institute{
INAF-Osservatorio Astronomico di Bologna, Via Ranzani 1, I-40127 Bologna, Italy
\and
INAF-Osservatorio Astronomico di Padova, Vicolo dell'Osservatorio 5, I-35122
 Padova, Italy
\and
Dept. of Physics and Astronomy, Macquarie University, Sydney, NSW, 2109 Australia 
\and
Monash Centre for Astrophysics, Monash University, School of Mathematical 
Sciences, Building 28, Clayton VIC 3800, Melbourne, Australia
\and
European Southern Observatory, Alonso de Cordova 3107, Vitacura, Santiago, Chile
\and
INAF-Osservatorio Astrofisico di Catania, Via S.Sofia 78, I-95123 Catania, Italy
\and
Dipartimento di Fisica e Astronomia, Universit\`a di Catania, Via S.Sofia 78, 
 I-95123 Catania, Italy 
  }

\date{}

\abstract{We present the abundance analysis of 82 red giant branch stars in the
dense, metal-poor globular cluster NGC~6093 (M~80), the largest sample of stars
analyzed in this way for this cluster. From high resolution UVES spectra of 14
stars and intermediate resolution GIRAFFE spectra for the other stars we derived
abundances of O, Na, Mg, Al, Si, Ca, Sc, Ti, V, Cr, Mn, Fe, Co, Ni, Cu, Zn, Y,
Zr, Ba, La, Ce, Pr, Nd, Sm, Eu. On our UVES metallicity scale the mean metal
abundance of M~80 is [Fe/H]$=-1.791\pm0.006\pm0.076$ ($\pm$statistical
$\pm$systematic error) with $\sigma=0.023$ (14 stars). M~80 shows star to star
variations in proton-capture elements, and the extension of the Na-O
anticorrelation perfectly fit the relations with (i) total cluster mass, (ii)
horizontal branch morphology, and (iii) cluster concentration previously found
by our group. The chemistry of multiple stellar populations in M~80 does not
look extreme. The cluster is also a typical representative of halo globular
clusters for what concerns the pattern of $\alpha-$capture and Fe-group elements.
However we found that a significant contribution from the $s-$process is
required to account for the distribution of neutron-capture elements. A minority
of stars in M~80 seem to exhibit slightly enhanced abundances of $s-$process 
species, compatible with those observed in M~22 and NGC~1851, although further
confirmation from larger samples is required.}

\maketitle

\section{Introduction}

The observational evidence of multiple stellar populations in Galactic globular
clusters (GCs) was already available since the late '70s, in the large
star-to-star abundance variations of light elements (C, N, O, Na, Mg, Al)
following patterns of correlations and anticorrelations  (see the reviews by
Smith 1987, Kraft 1994, Sneden 2000, Gratton, Sneden \& Carretta 2004). 
However, only in the late '80s/early '90s Denisenkov \& Denisenkova (1989) and
Langer et al. (1993) were able to explain the observed abundance variations with
the simultaneous action of the Ne-Na and  Mg-Al chains in the same temperature
layers where the ON part of the complete CNO cycle was at work. 
The
discovery of Na-O and Mg-Al anticorrelation among turn-off and subgiant (SGB)
stars first in NGC~6752 (Gratton et al. 2001) and then in other GCs (M~71,
Ram\'irez \& Cohen 2002; M~5, Ram\'irez \& Cohen 2003; 47~Tuc, Carretta et al.
2004) provided the key spectroscopic evidence of multiple populations. These
unevolved stars cannot either synthesise or transport these elements to the
surface, thus they must  be formed from gas polluted by ejecta of more massive
stars of an earlier stellar generation. Hence, ``observing an anticorrelation
among proton-capture elements" is simply a rewording for ``seeing multiple
stellar populations".

The advent of efficient, high-multiplexing spectrographs permitted, in the last
10 years, to collect precise abundances for $large$ samples of stars from high
resolution spectroscopy. We exploited FLAMES@VLT and designed a large, systematic
survey of RGB stars in Galactic GCs (for results, see e.g., Carretta et al.
2006, 2009a,b,c, 2014a,b,c, and references therein;  D'Orazi et al. 2010a, 2014;
Gratton et al. 2006, 2007). This investigation was recently extended to other
evolutionary phases like horizontal branch (HB), SGB and dwarf stars (e.g.,
D'Orazi et al. 2010b; Gratton et al. 2011, 2012a, 2014). Other, independent studies
concentrated mostly on the closest GCs like 47~Tuc (e.g. Cordero et  al. 2014,
Dobrovolskas et al. 2014), M~4 (e.g. D'Orazi \& Marino 2010, D'Orazi et al.
2013, Marino et al. 2008), NGC~6397 (e.g. Lind et al. 2008, 2009;  Gonz\'alez
Hern\'andez et al. 2009; Lovisi et al. 2012), NGC~6752 (e.g.  Yong et al. 2005,
2013;  Gruyters et al. 2014, Shen et al. 2010), and M~22 (Marino et al. 2011,
2012, 2013).  The literature is vast\footnote{We concentrate on spectroscopy in
the present paper. However, the improvements in precision photometry from ground
or space-based facilities produced colour-magnitude diagrams (CMDs) of GCs 
revealing more and more split SGBs, RGBs, and even multiple main sequences
(e.g., Grundahl et al. 1998, Han et al. 2009, Lee et al. 2009, Monelli et al.
2013, Milone et al. 2012,2013, Nardiello et al. 2015 and references therein).
The synergy between spectroscopy and photometry allows to uncover how absorption
features due to the involved light elements affect the colour indexes of split
sequences (e.g. Carretta et al. 2011a, Sbordone et al. 2011, Milone et al. 2012,
Cassisi et al. 2013).} and the reader is directed to the review by Gratton et
al. (2012b) and to the references in the more recent papers.

Our homogenous analysis of a large sample of GCs and large numer of stars in
each GC permits a quantitative approach (e.g. Carretta et al. 2009a,b), showing
that the Na-O anticorrelation differs from cluster to cluster in shape and
extent (measured by the interquartile range -IQR- of the [O/Na] ratio,
introduced by Carretta 2006). Adding also literature data, we see that the Na-O
anticorrelation stands out as the chief chemical signature of multiple stellar 
populations, so widespread that it may even be used to define a GC and that the
phenomenon is primarily driven by the cluster mass (Carretta et al. 2010a). 
However, our ignorance about the still elusive nature of first
generation polluters hampers our vision. Massive stars, either single, and fast
rotating: (FRMS: Decressin et al. 2007) or in close binary systems (de Mink et
al. 2009), intermediate-mass asymptotic giant branch (IM-AGB) and super-AGB
stars (Ventura et al. 2001, D'Ercole et al. 2012), or none of them (Bastian et
al. 2013), have been proposed as polluters. It is even possible that $both$ the
two commonly favoured candidates (FRMS and IM-AGB) were at work. In two GCs
where RGB stars were recently found clustered into discrete groups according to
their distinct composition (NGC~6752: Carretta et al. 2012; NGC~2808: Carretta
2014) the population with intermediate chemistry cannot be reproduced by mixing
unprocessed and heavily polluted matter from the primordial and the extremely
processed components, calling for at least two kind of polluters.

Our approach to these problems has been to gather an unprecedented large
statistics, both in term of number of GCs and of stars for which homogeneous
abundances were derived in each GC. We studied
24 massive GCs with different global parameters (metallicity, age, Galactic
population, HB morphology, concentration, etc.) and we present here the
abundance analysis of the last of them, NGC~6093 (M~80).

M~80 is a metal-poor ([Fe/H]$=-1.73$ dex, Cavallo et al. 2004), moderately
massive ($M_V=-8.23$ mag, Harris 1996, 2010 edition) cluster of Oosterhoff type
II (Kopacki 2013), one of the 30  densest GCs in the Milky Way. In the central
regions of M~80 Ferraro et al. (1999) identified over 300 blue straggler stars,
despite the relatively low inferred collision rate. Their inference was that
M~80 could be in an unusual dynamical state, on the edge of core collapse where
the density has recently become large enough to increase the
encounters involving primordial binaries, triggering a burst of blue straggler
formation (but see Heinke et al. 2003 for a different view, based on the
comparison with 47~Tuc and its possibly uncomplete census of blue stragglers).
The HB morphology is characterised by an extended blue tail (Ferraro et al.
1998), with the concentration of blue tail stars increasing toward the cluster
centre (Brocato et a;. 1998). M~80 presents a typical halo-type orbit,
with a short period and small size, never leaving the inner 3-4 kpc of the
Galaxy (Dinescu et al. 1999). 

The only chemical evidence of multiple populations comes from the study of ten
RGB stars by Cavallo et al. (2004) who found a mean [Al/Fe] ratio 
of +0.37 dex with a large spread of 0.43 dex (1 sigma) in aluminum abundance.
Recently Monelli et al. (2013) used private photometry from their SUMO project
to study the colour spread of the RGB.

The paper is organized as follows: in \S2 we present the basic data  for
this cluster, the selection of stars to be observed, and the observations
themselves. 
In \S3 we describe the analysis method and the results for abundances are 
presented in \S4, whereas they are discussed in \S5 in the context of multiple
populations in GCs. Finally, in \S6 we summarize our findings.

\section{Observations}\label{obs}
The photometric catalog for NGC~6093 is based on $BV$ data collected  at the
Wide-Field Imager at the 2.2-m ESO-MPI telescope on 9 July 1999. These data were
reduced by YM in a standard way (see Carretta et al. 2014a, Momany et al. 2004
for details) and the photometry is unpublished. 

\begin{figure}
\centering
\includegraphics[bb=20 150 330 715, clip, scale=0.825]{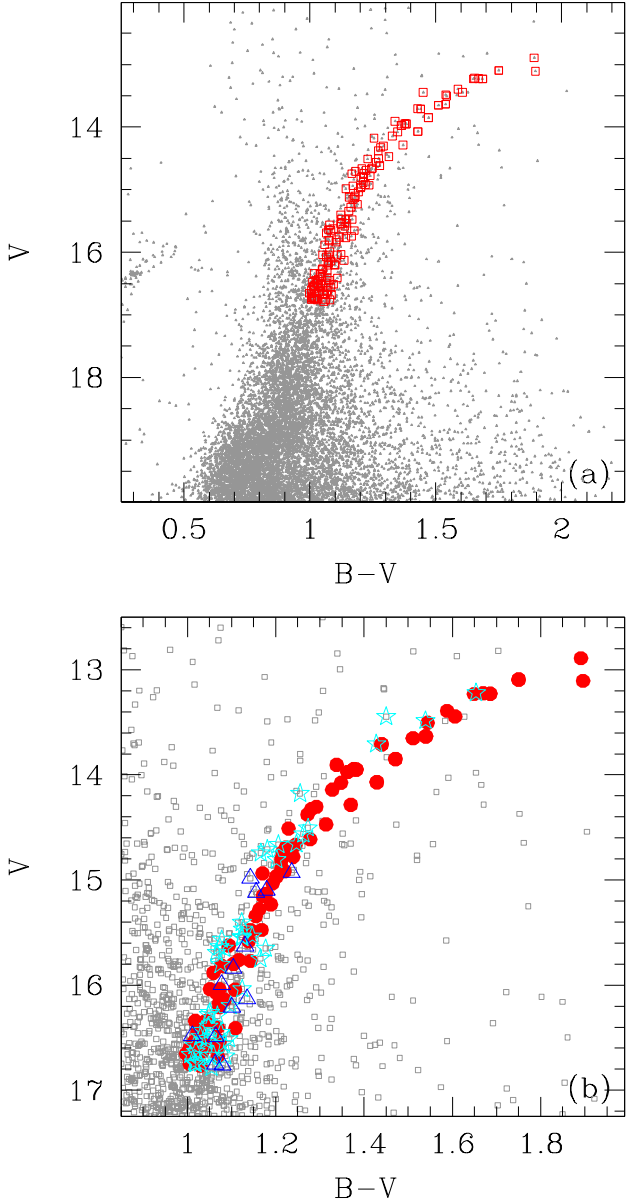}
\caption{(a) CMD of the cluster, with FLAMES targets indicated by red, open
symbols. (b) Enlargmenent of the RGB region, with member stars indicated by
filled red dots, non members on the basis of RV by cyan open stars, non members
on the basis of abundance by open blue triangles.}
\label{fig1}
\end{figure}

Following our standard procedure, we selected a pool of stars lying near the RGB
ridge line in the CMD and without close neighbours.  
The stars in our spectroscopic sample are indicated as large symbols in
Fig.~\ref{fig1}(a), with a zooming of the RGB region in
Fig.~\ref{fig1}(b), where we also indicate the membership status of our
targets (based on the analysis of their spectra).   NGC~6093 is heavily contaminated by field stars. The excision of
non-members was generally done via radial velocity (RV); however, in some cases,
we discarded stars on the basis of their metallicity (see next sections). 

\begin{figure}
\centering
\includegraphics[bb=20 150 288 706, clip, scale=0.825]{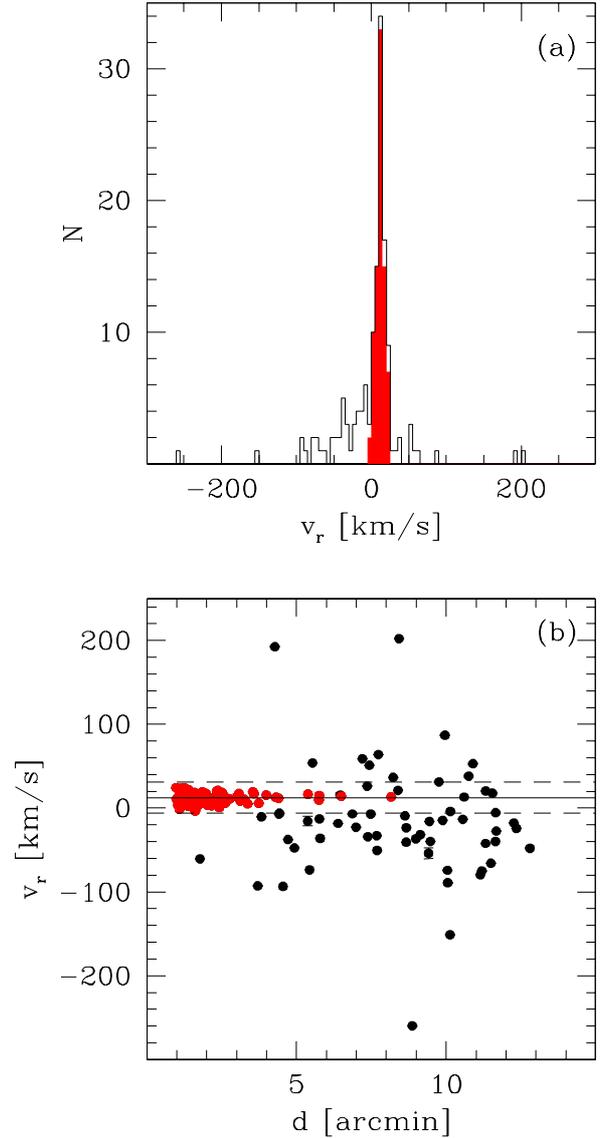}
\caption{(a) Histogram of RV values for all stars observed. (b) Plot of RV as
function of distance from the cluster centre; in red we indicate the member
stars. The lines show the average RV and the $\pm3\sigma$ limits.}
\label{fig2}
\end{figure}

We used FLAMES@VLT and the log of the observations is given in
Table~\ref{t:logobs}. We obtained three exposures with the high resolution
GIRAFFE grating and the setup HR11 covering the Na~{\sc i} 5682-88~\AA\ doublet
and three exposures with the setup HR13 including the [O~{\sc i}] forbidden
lines at 6300-63~\AA \ and another Na~{\sc i} doublet at 6154-60~\AA.  We
observed a total of 14 giants with the fibres feeding the UVES spectrograph (Red
Arm, with spectral range from 4800 to 6800~\AA\ and R=47,000). All the 14 stars 
turned out to be  members. We observed 78 member stars of M~80 with GIRAFFE; ten
are in common with UVES, so the grand total of our sample consists of 82 red
giant stars. 

We used the 1-D, wavelength calibrated spectra as reduced by the ESO personnel 
with the dedicated FLAMES pipelines. 
RVs for stars observed with the GIRAFFE spectrograph were
obtained using the  {\sc IRAF}\footnote{IRAF is  distributed by the National
Optical Astronomical Observatory, which are operated by the Association of
Universities for Research in Astronomy, under contract with the National Science
Foundation } task  {\sc FXCORR}, with appropriate templates, while those of the
stars observed with UVES were derived with the {\sc IRAF} task {\sc RVIDLINES}. 
The multiple exposures were then combined. 
The median values of the S/N ratio of the combined spectra are 125, 68, and 
78, for observations  with UVES and with the GIRAFFE HR11 and
HR13 setups, respectively.

Our optical $B, V$ photometric data were integrated with $K$ band magnitudes 
from the Point Source Catalogue of 2MASS (Skrutskie et al. 2006) to derive
atmospheric parameters as described in Section 3.

Coordinates, magnitudes, and heliocentric RVs are shown in Table~\ref{t:coo60}
(the full table is only available in electronic form at CDS).

\begin{table}
\centering
\caption{Log of FLAMES observations for NGC 6093.}
\begin{tabular}{cccccc}
\hline\hline
   Date         &     UT       & exp. & setup & seeing & airmass\\
  (yyyy-mm-dd)  & (hh:mm:ss)   & (s)  &         &  (")   &        \\ 
\hline
2009-09-08 &  00:28:42.650  &  2940    & HR11  & 0.89	&1.237  \\
2009-09-09 &  00:50:31.604  &  2940    & HR11  & 0.91	&1.343  \\
2009-08-06 &  00:16:09.457  &  2940    & HR11  & 0.95	&1.002  \\
2009-09-10 &  00:02:22.285  &  2940    & HR13  & 1.44	&1.177  \\
2009-09-12 &  23:40:18.289  &  2940    & HR13  & 0.86	&1.149  \\
2009-08-06 &  00:16:09.457  &  2940    & HR13  & 0.94	&1.330  \\
\hline
\end{tabular}
\label{t:logobs}
\end{table}

According to Harris (1996, 2010 edition), the metallicity of NGC~6093 is  $-1.75$
and the RV is $8.1\pm1.5$ km~s$^{-1}$ .
Our choice of member stars rests both on RV and, for marginal or dubious cases,
also on metallicity. Figure~\ref{fig2} shows the histogram of the heliocentric
RVs (in panel a) and the run of RVs with distance from the cluster centre (panel
b). We selected as candidate  cluster members 82 stars for which both RV and
metallicity were acceptable (see Table~\ref{t:coo60}). Not surprisingly, they
are concentrated at small distances from the cluster centre. For them we  find
an average velocity of  $\langle RV \rangle=11.9\pm0.7$~km~s$^{-1}$ ($rms=$ 6.1
km~s$^{-1}$), in  good agreement with the value reported by Harris (1996).

\subsection{Kinematics}

The spectroscopic dataset presented here constitutes the largest sample of
intermediate/high resolution spectra collected so far for NGC 6093, and can
therefore be used to study the kinematical properties of this cluster.

\begin{figure}
\centering
\includegraphics[scale=0.40]{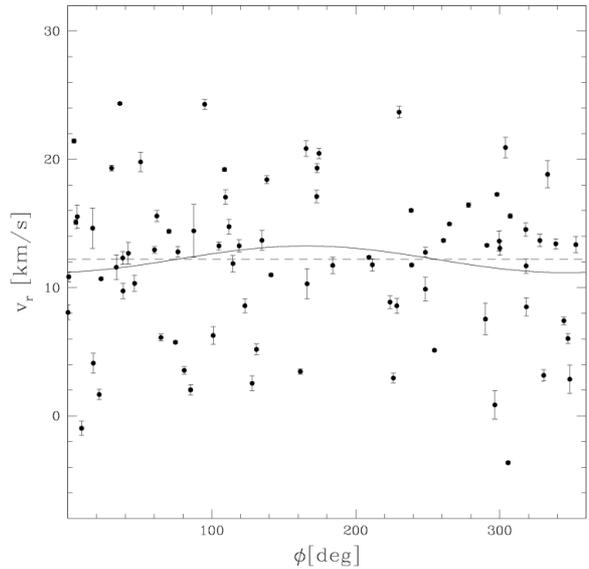}
\caption{Radial velocities of member stars in M~80 as a function of their
position angle.}
\label{f:rot}
\end{figure}

In Fig.~\ref{f:rot} the radial velocities of the member stars are plotted
against their position angle. To test the possible presence of systemic rotation
we calculated the difference between the average velocity of stars located
within and outside a $180^\circ$ interval centered around various position
angles, and compared it with a set of Monte Carlo extractions of the same number
of stars randomly distributed across the field of view. We found no significant
signature of rotation with a maximum difference lying at $\sim 1 \sigma$ from
the mean distribution of randomly extracted stars. The bestfit sinusoidal curve
overplotted to  Fig.~\ref{f:rot} has a maximum amplitude of
$A_{rot}=1.0\pm0.6~$km~s$^{-1}$, corresponding to a ratio
$A_{rot}/\sigma_{0}=0.09\pm0.06$ (see below). This value places NGC~6093 in the
group of non/slowly rotating GCs of Bellazzini et al. (2012). 

We then fitted the radial distribution of velocities with a set of single-mass
King (1966) and multi-mass King-Michie (Gunn and Griffin 1979) isotropic models.
For this purpose, the density profile has been derived by counting stars
brighter than the turn-off point (well above the limiting magnitude of the WFI 
photometry where severe incompleteness affects star counts) in
concentric annuli at different distance to the cluster center. The obtained 
density profile has been then matched with that provided by Noyola and Gebhardt
(2006) from integrated photometry on Hubble Space Telescope images to sample the
innermost 1.7\arcmin\ where the WFI spatial resolution does not allow a complete
sampling of stars. The total cluster $V$ magnitude, obtained by integrating the
observed profile, turns out to be $V=7.48\pm0.05$ corresponding to a total
luminosity of $log L/L_{\odot}=5.18\pm0.09$.
The velocity dispersion profile has been derived by dividing the sample in four
radial bins containing $\sim20$ stars each. 

\begin{figure*}
\centering
\includegraphics[bb=39 154 572 430, clip, scale=0.52]{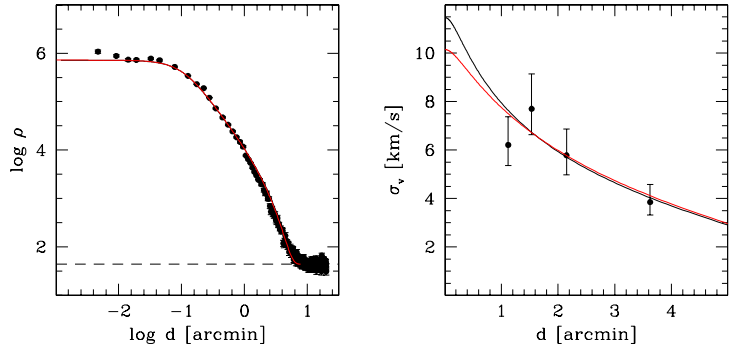}
\caption{Density and velocity dispersion profiles (left and right panel,
respectively) in M~80. The best-fit models are supersimposed (solid red line).}
\label{f:prof}
\end{figure*}

Multi-mass models has been constructed by assuming nine
mass bins and a mass function derived iteratively to match the observed star
counts in the deep ACS photometry of NGC~6093 from the {\it ACS globular 
clusters treasury project} (Sarajedini et al. 2007) corrected for completness 
(see Anderson et al. 2008). We adopted the same procedure described
in Sollima et al. (2012) to account for dark-remnants and binary heating.
The dynamical masses of the best-fit models have been
calculated using the maximum likelihood algorithm described by Pryor and Meylan
(1993) applied to our sample of radial velocities. The estimated central line of
sight (LOS) velocity dispersions and the corresponding dynamical masses are
$\sigma_{v,0}=11.4\pm0.9$~km s$^{-1}$ and  $\log{M/M_{\odot}}=5.50\pm0.07$ for the
single-mass model and $\sigma_{v,0}=10.1\pm0.8$~km s$^{-1}$ 
$\log{M/M_{\odot}}=5.56\pm0.07$ for the multi-mass model. The bestfit models are
overplotted to the density and velocity dispersion profiles in Fig.~\ref{f:prof}.
It is apparent that both models provide a good fit of both profiles. The
resulting M/L ratio is $2.1\pm0.6$, slightly larger than that predicted for a
mass-follow-light stellar system with the mass function slope estimated in the
literature ($\alpha=-1.36$; Paust et al. 2010), although the large associated
error prevents any firm conclusion.  This is in agreement with what found
by Sollima et al. (2012) for a sample of six Galactic GCs and interpreted as a
consequence of a high retention efficency  of dark remnants and/or to tidal
heating. In this regard, NGC~6093 follows an eccentric orbit (e=0.85) in a inner
region of the Milky Way (Dinescu et al. 1999) being therefore subject to an
intense tidal stress which can inflate its velocity dispersion even within its
half-mass (K\"upper et al. 2010).

\section{Atmospheric parameters, metallicity and abundance analysis}

In the abundance analysis we used mostly equivalent widths ($EW$s), measuread as
described in details in Bragaglia et al. (2001) using the package ROSA (Gratton
1988). The $EW$s measured on the GIRAFFE spectra were shifted to a system
defined by $EW$s from the high resolution UVES spectra. For this correction we
used 275 lines measured in the 10 stars observed with both instruments.

Abundances were derived from $EW$ analysis, using  the Kurucz (1993) grid of
solar-scaled LTE model atmospheres with the overshooting option switched off.

Effective temperatures T$_{\rm eff}$ were derived following our usual
two-steps procedure: first pass values were obtained from $V-K$ colours and the
Alonso et al. (1999, 2001) calibration, and these values were used as inputs to
derive the finally adopted T$_{\rm eff}$'s from a relation between 
T$_{\rm eff}(V-K)$ and the star magnitudes. Given the non negligible value of
the reddening  for M~80 ($E(B-V)=0.18$ mag, Harris 1996), we employed near
infrared $K$ magnitudes for this relation.

Surface gravities $\log g$ were obtained from bolometric corrections (from
Alonso et al.), the adopted effective temperatures, reddening and distance
modulus from Harris (1996), and assuming  masses of 0.85 M$_\odot$ and  $M_{\rm
bol,\odot} = 4.75$ as bolometric magnitude for the Sun.

Finally, we obtained values of the microturbulent velocity $v_t$ by eliminating
trends of the abundances from Fe~{\sc i} lines with the expected line strength 
(see Magain 1984). The final adopted
atmospheric parameters are listed with the derived Fe abundances of individual
stars in Tab.~\ref{t:atmpar60}.

Errors in atmospheric parameters and their impact on the derived abundances are
estimated as usual (see Carretta et al. 2009a,b and the Appendix to the present
paper). Internal (star to star) errors in temperature, gravity, model abundance
and $v_t$ are 4 K, 0.04 dex, 0.02 dex, and 0.10 km~s$^{-1}$,
respectively\footnote{The internal error in $v_t$
increases to 0.36 km~s$^{-1}$ for stars with GIRAFFE spectra.}. 

\begin{figure}
\centering 
\includegraphics[bb=78 171 460 691, clip,scale=0.52]{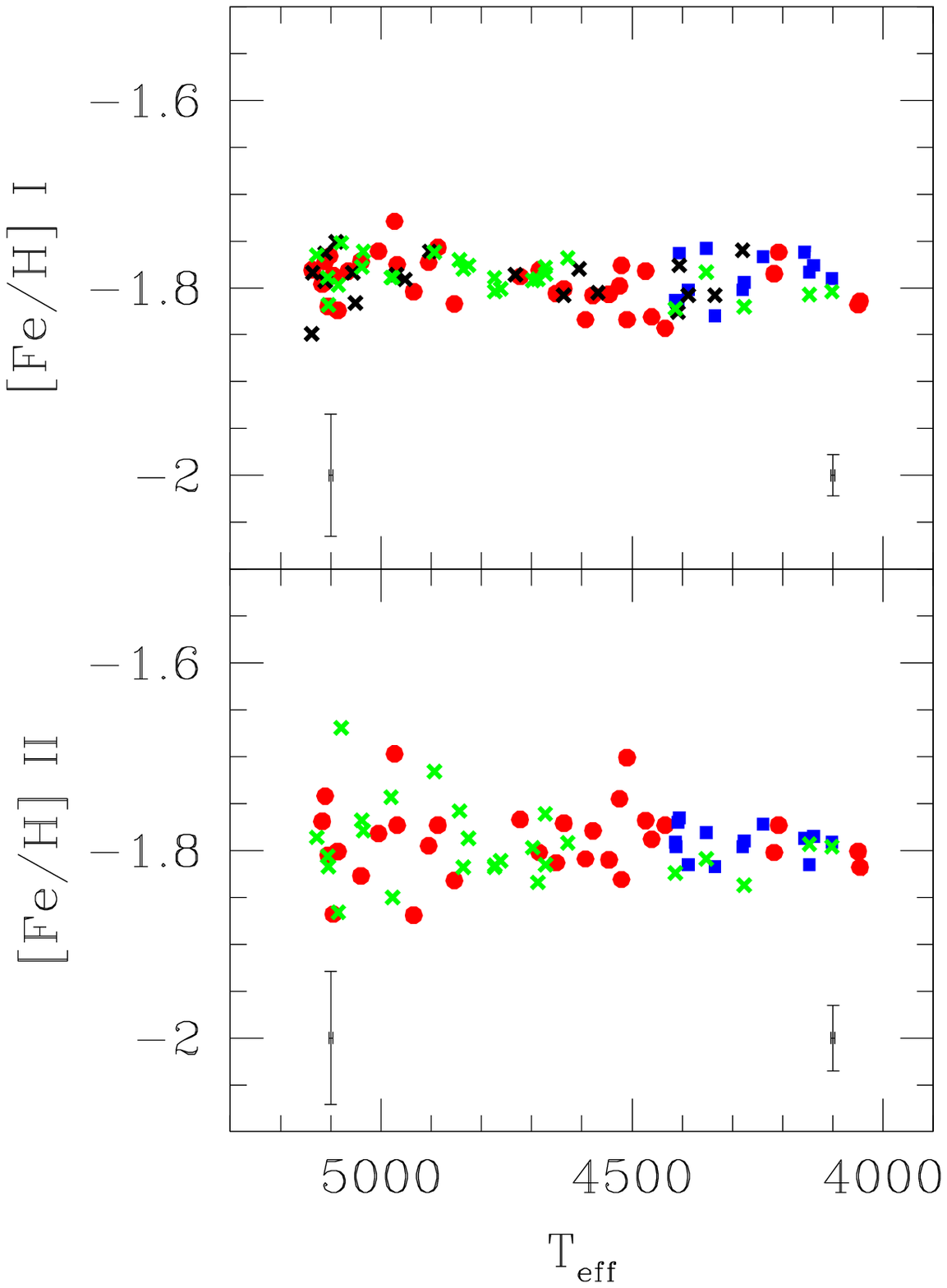}
\caption{Abundance ratios [Fe/H] {\sc i} (upper panel) and [Fe/H] {\sc ii}
(lower panel) as a function of T$_{\rm eff}$ for all member stars analysed in
M~80 with UVES (blue squares) and GIRAFFE spectra (red circles: setups HR11 and
HR13; black crosses: HR11 only; green crosses: HR13 only). Error bars on the
right and on the left side are star-to-star errors for targets observed with
UVES and GIRAFFE, respectively.}
\label{f:feteff60}
\end{figure}

As in the vast majority of GCs, the metallicity is very homogeneous in M~80. We
find on average [Fe/H]~{\sc i}$=-1.791\pm0.006\pm0.076$ dex ($\sigma=0.023$
dex)  from the 14 stars observed with UVES and 
[Fe/H]~{\sc i}$=-1.791\pm0.003\pm0.070$ dex ($\sigma=0.023$ dex) from 78 stars 
with GIRAFFE spectra (where the first and second error bars refer to the
statistical and systematic errors, respectively). The agreement with average
abundances derived from singly ionized Fe lines is very good: 
[Fe/H]~{\sc ii}$=-1.791$ dex $(\sigma=0.016$ dex, 14 stars with UVES) and 
[Fe/H]~{\sc ii}$=-1.792$ dex $(\sigma=0.040$ dex, 56 stars with GIRAFFE). The 
derived abundances show no
trend as a function of the temperature (Fig.~\ref{f:feteff60}) and are in good
agreement, within the uncertainties, with the average value found by Cavallo et
al. (2004) from 10 giants.

We measured the abundances of 12 species (beside Fe: O, Na, Mg, Si, Ca, Ti, 
Sc, V, Cr, Ni, and Ba) from both UVES and GIRAFFE
spectra. Additionally, from the UVES spectra with large spectral coverage we 
also obtained abundances of Al, Ti (from both Ti~{\sc i} and Ti~{\sc ii} lines),
Cr (both from Cr~{\sc i} and Cr~{\sc ii} lines), Mn, Co, Cu, Zn, 
Zr (both from Zr~{\sc i} and Zr~{\sc ii} lines), La, Ce, Pr, Nd, Sm, and Eu.
Adopted line list and solar reference abundances are from Gratton et al. (2003).
We did apply NLTE corrections to the Na abundances following the prescriptions
by Gratton et al. (1999). We applied abundance corrections for Sc, V, Mn, and 
Co to account for the  hyperfine structure (references in Gratton et al. 2003).

\section{Results}

\begin{figure*}
\centering 
\includegraphics[scale=0.52]{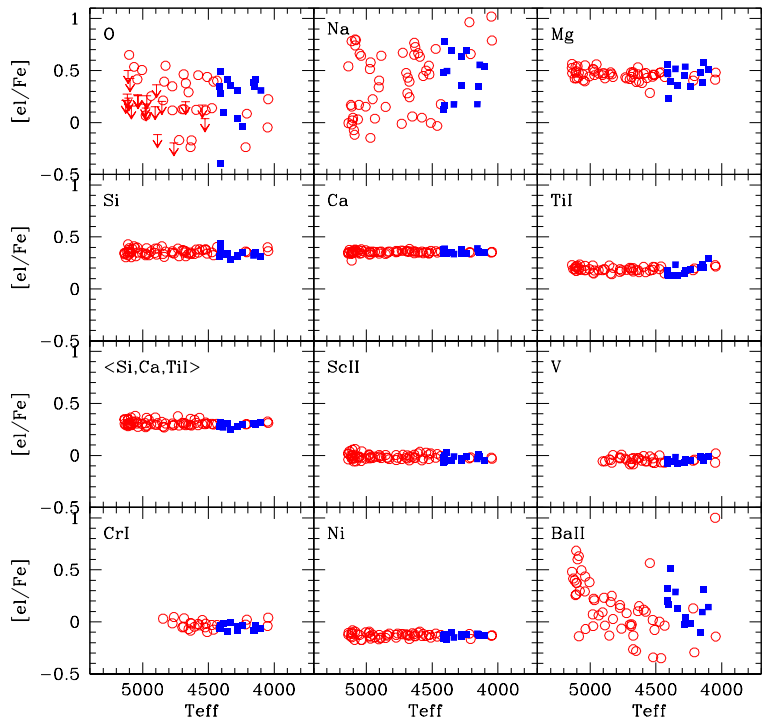}
\caption{Abundance ratios [el/Fe] as a function of T$_{\rm eff}$ for all 
stars in our sample. Filled squares and open circles indicate stars with UVES 
and GIRAFFE spectra, respectively. Internal error bars are listed in
Tab.~\ref{t:sensitivityu60} and Tab.~\ref{t:sensitivitym60} in the Appendix.}
\label{f:eleteff60}
\end{figure*}

The average values of all measured elements with their $r.m.s.$
scatter are listed in  Tab.~\ref{t:meanabu60}. Abundances of 
proton-capture, $\alpha-$capture, Fe-group and neutron-capture elements are
given for individual stars in Tab.~\ref{t:proton60}, Tab.~\ref{t:alpha60},
Tab.~\ref{t:fegroup60}, Tab~\ref{t:neutron60} and Tab.~\ref{t:ba60}.
Abundance ratios as a function of temperature are shown in
Fig.~\ref{f:eleteff60}. In this figure we also plot the average [$\alpha$/Fe]
values, where the mean includes [Si/Fe], [Ca/Fe], and [Ti/Fe]~{\sc i}.

\setcounter{table}{3}
\begin{table}
\centering
\caption{Mean abundances from UVES and GIRAFFE }
\begin{tabular}{lcc}
\hline
                     &               &               \\
Element              & UVES	     & GIRAFFE       \\
                     &n~~   avg~~  $rms$ &n~~	avg~~  $rms$ \\        
\hline
$[$O/Fe$]${\sc i}    &14   +0.24 0.24 &54   +0.23 0.21  \\
$[$Na/Fe$]${\sc i}   &14   +0.44 0.22 &63   +0.40 0.30  \\
$[$Mg/Fe$]${\sc i}   &14   +0.45 0.10 &66   +0.46 0.05  \\
$[$Al/Fe$]${\sc i}   &14   +0.48 0.34 & 	        \\
$[$Si/Fe$]${\sc i}   &14   +0.34 0.04 &75   +0.36 0.03  \\
$[$Ca/Fe$]${\sc i}   &14   +0.36 0.02 &78   +0.36 0.02  \\
$[$Sc/Fe$]${\sc ii}  &14 $-$0.03 0.03 &78 $-$0.01 0.03  \\
$[$Ti/Fe$]${\sc i}   &14   +0.19 0.05 &64   +0.19 0.02  \\
$[$Ti/Fe$]${\sc ii}  &14   +0.18 0.02 & 	        \\
$[$V/Fe$]${\sc i}    &14 $-$0.04 0.02 &45 $-$0.04 0.03  \\
$[$Cr/Fe$]${\sc i}   &14 $-$0.05 0.03 &34 $-$0.03 0.04  \\
$[$Cr/Fe$]${\sc ii}  &14 $-$0.01 0.03 & 	        \\
$[$Mn/Fe$]${\sc i}   &14 $-$0.48 0.02 & 	        \\
$[$Fe/H$]${\sc i}    &14 $-$1.79 0.02 &78 $-$1.79 0.02  \\
$[$Fe/H$]${\sc ii}   &14 $-$1.79 0.02 &56 $-$1.79 0.04  \\
$[$Co/Fe$]${\sc i}   &14 $-$0.22 0.03 &                 \\
$[$Ni/Fe$]${\sc i}   &14 $-$0.13 0.02 &69 $-$0.13 0.02  \\
$[$Cu/Fe$]${\sc i}   &14 $-$0.52 0.04 & 	        \\
$[$Zn/Fe$]${\sc i}   &14 $-$0.02 0.06 & 	        \\  
$[$Y/Fe$]${\sc ii}   &14 $-$0.07 0.04 & 	        \\ 
$[$Zr/Fe$]${\sc i}   & 8 $-$0.03 0.05 & 	        \\ 
$[$Zr/Fe$]${\sc ii}  &14   +0.00 0.06 & 	        \\ 
$[$Ba/Fe$]${\sc ii}  &14   +0.16 0.16 &59   +0.12 0.28  \\ 
$[$La/Fe$]${\sc ii}  &14   +0.28 0.11 & 	        \\ 
$[$Ce/Fe$]${\sc ii}  &14   +0.18 0.14 & 	        \\ 
$[$Pr/Fe$]${\sc ii}  &14   +0.24 0.09 & 	        \\ 
$[$Nd/Fe$]${\sc ii}  &14   +0.22 0.08 & 	        \\ 
$[$Sm/Fe$]${\sc ii}  &14   +0.40 0.08 & 	        \\ 
$[$Eu/Fe$]${\sc ii}  &14   +0.51 0.02 & 	        \\ 
\hline
\end{tabular}
\label{t:meanabu60}
\end{table}

\subsection{Proton-capture elements}

The Na~{\sc i} lines at 6154-60~\AA\ are weak in metal-poor, relatively warm 
stars, as discussed in Carretta et al. (2009a).
The possibility that they are
measured only when noise spuriously enhances their strength prompted us to adopt
an empirical parameter ($T_{\rm eff}/100$ K$-10 \times$[Fe/H]) to set rejection
criteria in these cases. We found that 12 stars in M~80 (mostly warmer than
$\sim4800$ K and with [Na/Fe]$\geq 0.5$ dex) with only one or both Na lines
from the weak doublet had the parameter exceeding the value of 65 fixed in
Carretta et al. (2009a). For sake of homogeneity, in these stars we dropped the
Na abundance although all these objects have a measured O abundance that would
put them reasonably well on the Na-O anticorrelation. After this culling of our
sample, we ended with O abundances for 63 RGB stars (44 actual detections and 19
upper limits) and Na abundances for 67 stars. The number of giants with
both O and Na abundances measured in M~80 is 50.

\begin{figure}
\centering 
\includegraphics[scale=0.40]{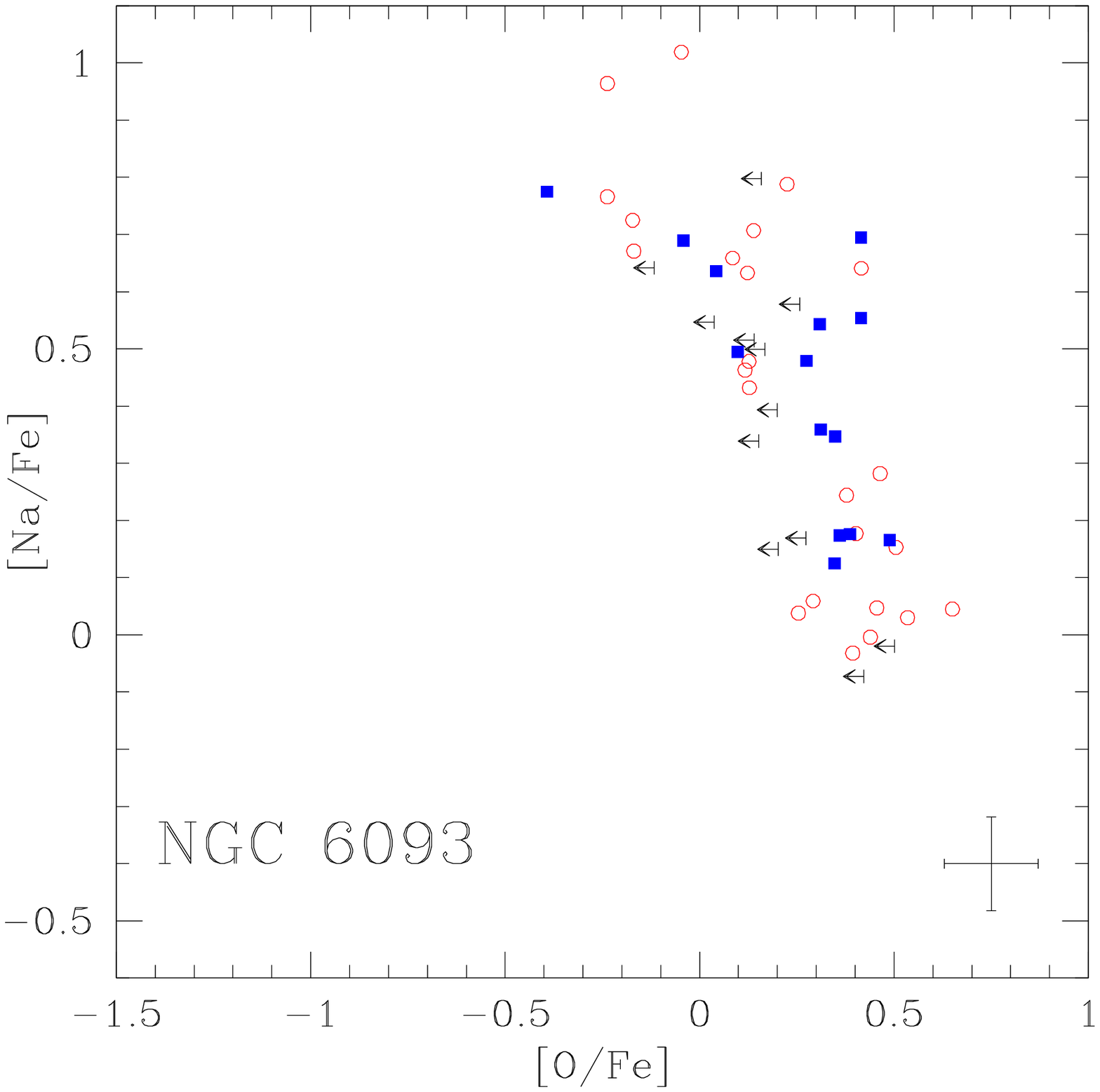}
\caption{The anticorrelation between Na and O abundances in M~80. Empty circles
indicate stars observed with GIRAFFE, filled squares with UVES. Upper limits in
O are indicated with arrows. The error bars represent internal errors (see
Appendix).}
\label{f:m60antiu}
\end{figure}

The spread in [O/Fe] and [Na/Fe] ratios in M~80 is significant, about 1 dex for
both elements. The resulting Na-O anticorrelation (Fig.~\ref{f:m60antiu})
is of moderate extension, reaching down [O/Fe]$\sim -0.5$ dex. This
chemical signature of multiple stellar populations presents the usual features:
a minority of stars with typical composition from core-collapse supernovae (high
O, low Na), separated more or less clearly at [Na/Fe]$\sim +0.3$ dex from the
bulk of stars with increasing Na and decreasing O abundances. Finally, a few
stars show signatures of extreme processing: very low O and high Na levels.

\begin{figure}
\centering 
\includegraphics[scale=0.40]{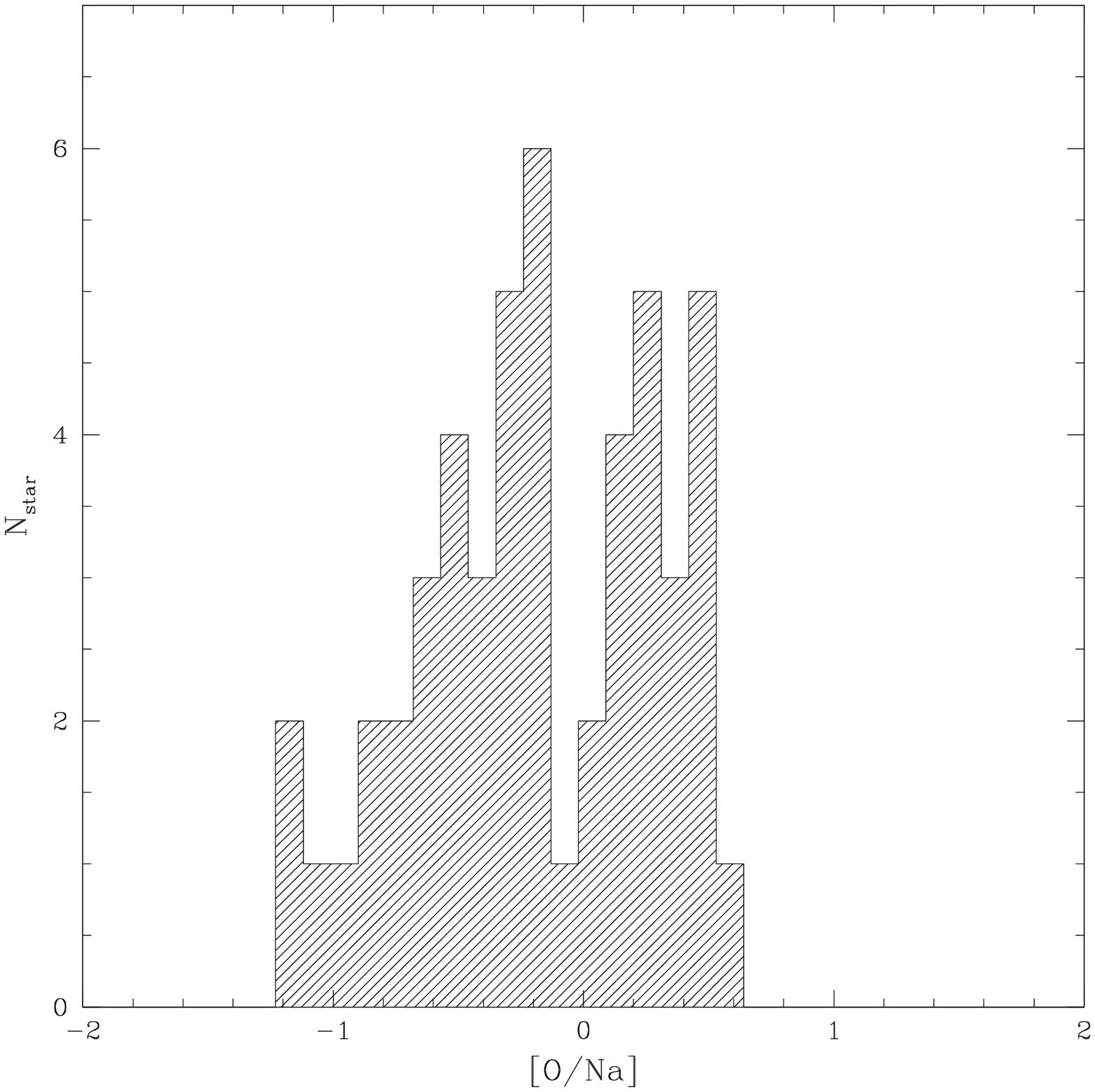}
\caption{The distribution of the [O/Na] ratios in M~80.}
\label{f:histoona}
\end{figure}

The different components are clearly apparent in the distribution of [O/Na]
ratios, shown in Fig.~\ref{f:histoona}: a dip between 0.0 and $-0.1$ dex separates
the two main populations, followed by a short tail at very low [O/Na] values.

Following the criteria defined in Carretta et al. (2009a) we can identify also
in M~80 a fraction of stars with primordial (P) composition, and two components
of second generation with intermediate (I) and extreme (E) chemistry processed
through hot H-burning. In M~80 the fractions we found are P$=36\pm8$\%
I$=56\pm11$\%, and E$=8\pm4$\%, respectively. We caution that due to the
limitations of the mechanical positioning of fibres our sample is composed of
stars located outside two half mass radii from cluster center. Since the global 
population fraction can be best characterized from stars between 1 and 2 half 
mass radii, first generation stars could be overrepresented in the observed 
sample, as pointed out by the referee. However, we note that the P, I, and E
fractions in M~80 are not exceptional with respect to the values found in other
GCs with a large variety of concentration, mass and dynamical age (see Carretta
et al. 2009a, their Table~5).

\begin{figure*}
\centering 
\includegraphics[scale=0.52]{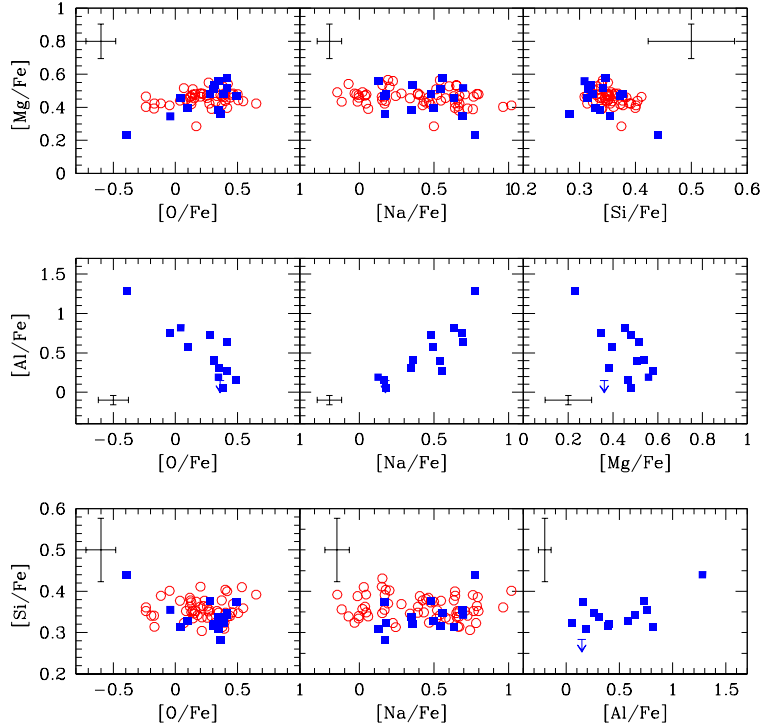}
\caption{Relations among the abundances of proton-capture elements in red giants
of M~80. Filled squares and empty circles indicate stars observed with UVES and
GIRAFFE, respectively. Internal error bars are also shown.}
\label{f:proton60}
\end{figure*}

The relations among all the abundances available for the light  elements O, Na,
Mg, Al, Si in M~80 are shown in Fig.~\ref{f:proton60}. Al was only available for
the 14 stars observed with UVES (one upper limit, 13 detections). Our average
value (+0.48 dex, $\sigma=0.34$ dex) is in good agreement with that from 10
stars in Cavallo et al. (2004: +0.37 dex, $\sigma=0.48$ dex). The [Al/Fe] ratios
present well defined anticorrelations with species depleted in proton-capture
reactions (O, and Mg) and a neat correlation with Na abundances (middle panels
of Fig.~\ref{f:proton60}). 

The other elements involved in hot H-burning (Mg, Si) do not seem to be
particularly touched by nuclear processing in now extinct first generation 
stars. The only exception is star 13987, observed with UVES, which shows the
lowest Mg and O abundances and, accordingly, the highest Al, Si (but not Na)
abundances in our sample. It is this object which drives the correlation
between Mg and O observed in our sample of stars with UVES spectra (top-left
panel in Fig.~\ref{f:proton60}) as well as the correlation Si-Al and the Mg-Si
anticorrelation (top-right panel and bottom right-panel, respectively, in
Fig.~\ref{f:proton60}).
We conclude (with a cautionary word about the limited size of our UVES
sample) that in M~80 the nuclear processing at extremely hot temperatures
affected only a trace fraction of stars, at odds with what occurred in other GCs,
e.g. NGC~4833 (Carretta et al. 2014b), a cluster with a very similar (present-day) 
total mass.

\subsection{$\alpha$ and Fe-group elements}

The run of $\alpha-$elements as a function of the temperature is shown in
Fig.~\ref{f:eleteff60}. Apart for a small increase in the scatter for Mg, the
other elements of this group (Si, Ca, Ti) are very homogeneous in M~80, with
dispersion totally explained by uncertainties in the analysis. In 
Fig.~\ref{f:eleteff60} we also plot the mean [$\alpha$/Fe] ratio, obtained 
from the average of [Si/Fe], [Ca/Fe], and [Ti/Fe]~{\sc i}: $+0.308\pm0.003$ dex
($\sigma=0.026$ dex, 82 stars). Including also Mg in the mean the value would
have been $+0.342\pm0.004$ ($\sigma=0.032$ dex, 82 stars), another evidence that
only a few stars in M~80 were polluted by material processed at the high
temperatures required for a significant depletion of Mg.

The elements of the Fe-group do not show any intrinsic scatter in M~80, and
usually track iron, apart from Mn and Cu (only measured in stars with UVES
spectra), which present the usual underabundance typical of metal-poor GCs (see
Gratton et al. 2004). 

In cases where we measured both neutral and singly ionized transitions for a
species (Fe, Ti, Cr), there is an excellent agreement between the abundances. 

\subsection{Neutron-capture elements}

We derived the abundances of a few neutron-capture elements, mainly for stars
with UVES spectra, apart from Ba, for which a transition was measurable also in
stars with GIRAFFE spectra and HR13 setup.

Concerning Ba, the lines available are strong and more sensitive to velocity
fields than to the abundances. To avoid the ensuing trends as a function of the
$v_t$ (whose values are derived from a plaethora of weaker Fe lines) we adopted
the procedure followed in Carretta et al. (2013, 2014a). Abundances of Ba were
obtained using a common [Fe/H] for all stars (equal to the mean value -1.79 dex)
and values of $v_t$ provided by the relation found by Worley et al. (2013).
The abundances from $EW$s of all three Ba lines available in the UVES spectra 
were checked with synthetic spectra employing the line
lists by D'Orazi et al. (2012), finding good agreement between the two methods.
The results are shown in the last panel of Fig.~\ref{f:eleteff60} as a function
of temperature: most of the observed spread is accounted for by uncertainties
derived from the analysis (see error tables in the Appendix). The abundances of
Ba are not correlated with any of the proton-capture elements.

\begin{figure}
\centering 
\includegraphics[scale=0.40]{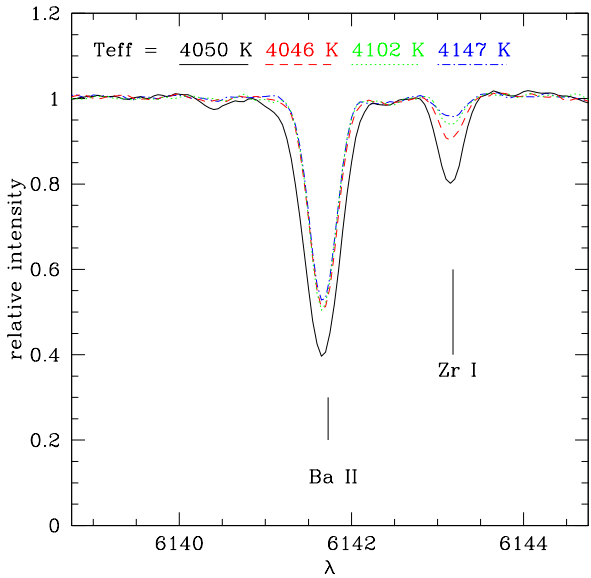}
\caption{The GIRAFFE spectrum of star 16162 (T$_{\rm eff}=4050$ K,
[Ba/Fe]=+1.001 dex) in the region of the Ba~{\sc ii} 6141~\AA\ line (solid line)
compared to stars with similar temperatures: 14224 (dashed line), 17886 (dotted
line), and 17874 (dot dashed line).}
\label{f:m60barich55}
\end{figure}

We note that star 16162 (with T$_{\rm eff}=4050$ K) stands out with its
[Ba/Fe]=1.001 dex, the highest in all our sample. This star has also the highest
Na abundance and it is characterized by the reddest $B-V$ colour in our dataset
of stars. In Fig.~\ref{f:m60barich55} the spectrum of this Ba-rich star is
compared in the region of the Ba~{\sc ii} 6141~\AA\ line to the spectra of the
three stars with most similar temperatures (and much lower Ba abundances, see
Tab.~\ref{t:ba60}). The spectral lines of star 16162 show a clear broadening
with respect to the other stars. We conclude that this object likely presents
hints of a phenomenon of mass transfer from a binary companion, although its RV
does not show particular differences with respect to other stars.
Anyway, by eliminating this star the average from GIRAFFE spectra changes only a
little, becoming [Ba/Fe]$=+0.11\pm 0.04$ dex ($\sigma=0.25$ dex, 58 stars).

The results for Y~{\sc ii} were validated through spectral synthesis of the 
strong line at 4883~\AA, using a line list from D'Orazi et al. (2013).  La and
Ce abundances were obtained from the lines at 6390~\AA\ and at 5274~\AA, using
spectrum synthesis and EWs, respectively, whereas Pr (4 lines) and Sm (2 lines)
were analyzed by using  atomic parameters from Sneden et al. (2003) and Koch and
McWilliam (2014). Abundances of Nd were derived as in Carretta et al. (2011b).

\section{Discussion: M~80 in context}

\subsection{Multiple stellar populations in M~80}

The spectroscopic observations in M~80 complete the core sample of our FLAMES
survey of the Na-O anticorrelation in Galactic globular clusters. We have
studied 24 GCs (about 15\% of those listed in the Harris 1996 catalogue): the
range of main parameters covered by our sample is reported in
Tab.~\ref{t:tabrange}, together with the values for M~80.

\setcounter{table}{9}
\begin{table}
\caption[]{Summary of the global parameters sampled in GCs of our FLAMES survey}
\begin{tabular}{lrrr}
\hline
parameter           	  & min  & max &NGC 6093\\
\hline        
$M_V$ (mag)         	  & -6.64    & -9.98	& -8.23 \\
                    	  & NGC 6397 & NGC 6715 &       \\
                          &          &          &       \\
R$_{GC}$ (kpc)      	  & 3.1      & 18.9	& 3.8   \\
                    	  & NGC 6388 & NGC 6715 &       \\
                          &          &          &       \\
$$[Fe/H]$$ (dex)    	  & -2.34    & -0.43	& -1.79 \\
                    	  & NGC 7099 & NGC 6441 &       \\
                          &          &          &       \\
c                   	  & 0.93     & 2.29	& 1.68  \\				
                    	  & NGC 6809 & NGC 7078 &       \\
                          &          &          &       \\
$\log t_c$ (years)  	  & 4.94     & 8.99	& 7.78  \\
                    	  & NGC 6397 & NGC 288  &       \\
                          &          &          &       \\
$\rho_0$ (L$_\odot/pc^3$) & 1.78     & 5.76     & 4.79  \\
                          & NGC 0288 & NGC 6397	&       \\
                          &          &          &       \\
r$_h$ (pc)                & 1.56     & 5.68     & 1.89  \\
                          & NGC 6397 & NGC 0288 &       \\
                          &          &          &       \\
r$_t$ (pc)                & 10.43    & 90.02    & 38.63 \\
                          & NGC 6838 & NGC 4590 &       \\
                          &          &          &       \\
HB index                  & -1.00    & 1.00     & 0.93  \\
                          & NGC 6388 & NGC 6752 &       \\
                          &          &          &       \\
rel. age                  & 0.84     & 1.06     & 1.03  \\
                          & NGC 0362 & NGC 7099 &       \\
                          &          &          &       \\
IQR[O/Na]                 & 0.257    & 1.169    & 0.784 \\
                          & NGC 6838 & NGC 6715 &       \\						    
                          &          &          &       \\			       
\hline
\end{tabular}
\begin{list}{}{}
\item[] - $M_V$, R$_{GC}$, c, $\log t_c$, $\rho_0$ from Harris (1996, 2010 edition).
\item[] - [Fe/H] from Carretta et al. (2009c,2010b,2011b,2013,2014b)
\item[] - c: excluding post-core collapse GCs (NGC 6397,6752,7099), where c=2.5
\item[] - r$_h$, r$_t$, HB index: Mackey \& Van den Bergh (2005)
\item[] - relative age: Carretta et al. (2010a)
\item[] - IQR[O/Na]: our FLAMES survey
\end{list}
\label{t:tabrange}
\end{table}

\begin{figure}
\centering
\includegraphics[bb=121 155 418 708, clip, scale=0.52]{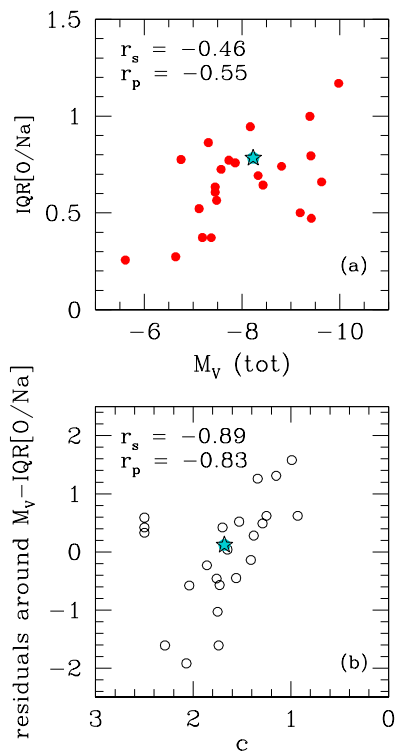}
\caption{Upper panel: correlation between total absolute magnitude and the
interquartile range of the [O/Na] ratio for GC in our FLAMES survey. Lower
panel: relation between cluster concentration and the residual around the best
fit relation IQR[O/Na]-$M_V$(tot). M~80 is indicated by a filled star symbol. In
each panel the Spearman rank correlation coefficient $r_s$ and the
Pearson linear correlation coefficient $r_p$ are listed.}
\label{f:mviqrres60}
\end{figure}

M~80 is a typical inner halo GC (Carretta et al. 2010a), and also the chemistry
of its multiple stellar populations appears to be normal. The extension of the
Na-O anticorrelation, the most prominent signature of multiple stellar
generations, is quantified by a IQR[O/Na]=0.784. This value locates M~80 right
in the middle of the main relations defined by other GCs in our core sample,
like the correlation between extent of the anticorrelation and total cluster
mass (upper panel of Fig~\ref{f:mviqrres60}). 

In Carretta et al. (2014b) we discovered a tight relation (once the post-core
collapse GCs are excluded) linking the cluster concentration and the residuals
around the relation between $M_V$ and IQR[O/Na]. More concentrated clusters seem
to develop a less extended Na-O anticorrelation with respect to more loose GCs.
The anticorrelation of M~80 is in very good agreement with this trend
(Fig.~\ref{f:mviqrres60}, bottom panel).

\begin{figure}
\centering 
\includegraphics[scale=0.40]{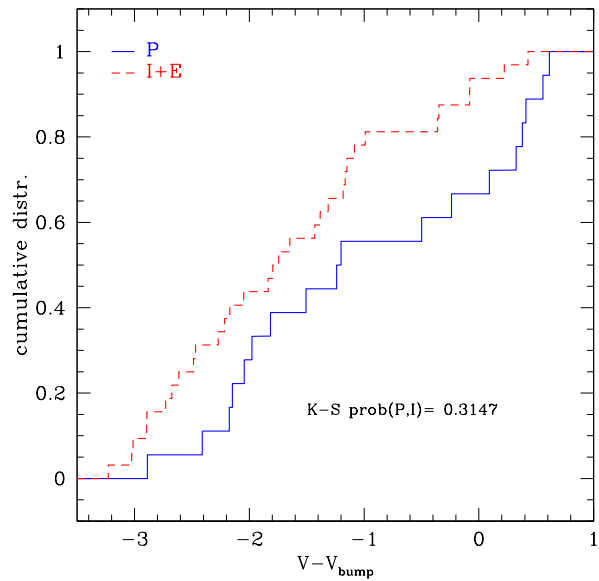}
\caption{Cumulative distribution of the difference in magnitude between
individual $V$ and $V_{bump}=16.12$ mag for all stars of the first generation
(solid line) and of the second generation (dashed line) with measured Na
abundances.}
\label{f:bumpcumu}
\end{figure}

In M~80 we were able to observe a number of stars below the bump level on the
luminosity function of the RGB ($V=16.12\pm0.03$ mag, Zoccali et al. 1999). As
predicted by Salaris et al. (2006), the luminosity of the RGB bump is sensitive
to the He abundances, being brighter for a population where a He-enhanced
component is present. This prediction has been since a long time observationally
confirmed both with spectroscopy (using the component with Na enhancement in GCs
as a proxy, see Carretta et al. 2007a, Bragaglia et al. 2010) and with
photometry (e.g. Nataf et al. 2011). In Fig.~\ref{f:bumpcumu} we show the
cumulative distribution of the magnitude differences $V-V_{bump}$ for all stars
in M~80 with a measured Na abundance, separating the first generation stars (P
component) from second generation stars (I+E components).

Although a Kolmogorov-Smirnov test formally does not allow us to reject the null
hypothesis that the two subsamples are extracted from the same parent
distribution, it is clear that regardless of adopted magnitude cut the second 
generation stars are on average located at slightly brighter luminosities than
the first generation giants, as they should be according to their He-enhanced
abundances.

In our present sample of giants in M~80 we did not find a statistically 
significant difference in radial distribution. However, taking into account also
the physical limitations imposed by the fibres positioning tool, a more suitable
approach to study the radial concentration of different stellar generations is
offered by photometry of large samples over an extended area on the cluster.
The combination  $(U-B)-(B-I)$, which Monelli et al. (2013) call $c_{U,B,I}$,
is a good way to study the separation of stars according to different populations in
a GC, sice it does not (significantly) depend on reddening
and distance modulus. 
There are 16 objects, including M~80, in common between the SUMO sample and our
FLAMES survey. Unfortunately the photometry of individual stars is still
unpublished, but Monelli et al. provide the width $W_{RGB}$ of the RGB in
$c_{U,B,I}$  that can be used to better investigate the relation between
spectroscopic and photometric properties of multiple stellar populations in M~80
and in other GCs.

For M~80 they derive a value $W_{RGB}=0.14$, which is the same width observed
for NGC~288 and M~4 (NGC~6121). This occurrence is clearly at odds with the
behaviour of these GCs concerning the Na-O anticorrelation, as shown in the
upper panels of Fig.~\ref{f:naohb}, where we plot the [Na/Fe] and [O/Fe] ratios
of each cluster from our strictly homogeneous FLAMES survey.
 
\begin{figure}
\centering
\includegraphics[scale=0.40]{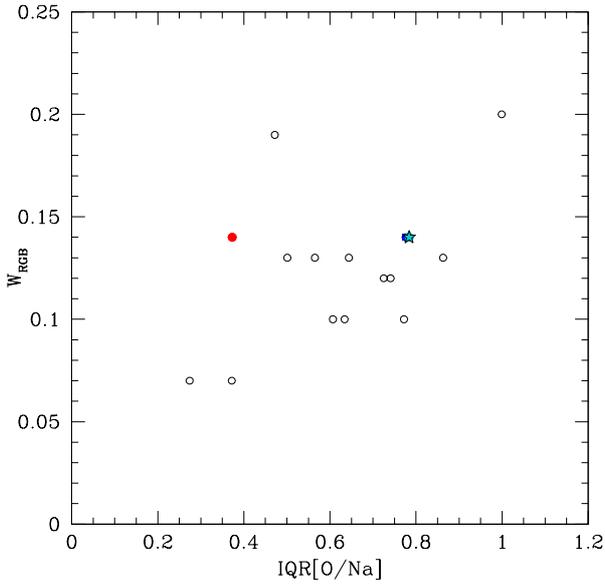}
\caption{The width of the RGB from the SUMO survey ($W_{RGB}$) as a function of
the interquartile range IQR[O/Na] from our FLAMES survey for the 16 GCs analyzed
in both samples. M~80 is indicated by a star symbol, under which NGC~288 (blue
filled square) is almost hidden. M~4 is indicated by a filled red circle.}
\label{f:sumo}
\end{figure}

In this figure we indicate with a vertical line the lowest border of the O
abundance measured in M~80, and superimposed this border to the other two plots.
The Na-O anticorrelation derived for M~4 (with IQR[O/Na]=0.373)\footnote{We used
the values derived from our homogeneous work, but other independent studies also
support the evidence that the Na-O anticorrelation is not very extended in M~4
(see e.g. Marino et al. 2008).} is clearly  much shorter than those observed in
both M~80 and NGC~288 (IQR=0.784 and 0.776, respectively).
Therefore it is not a surprise that $W_{RGB}$ and IQR[O/Na] are not well
correlated (see Fig.~\ref{f:sumo}). 
The Pearson's linear correlation coefficient is $r=0.46$ which with 16 GCs
indicates that the correlation is significant only at a level of confidence
between 90 and 95\%.

These three GCs are however different in other relevant global parameters: the
total metallicity decreases going from M~4 to NGC~288 and to M~80 
([Fe/H$=-1.17, -1.30, -1.79$ dex, respectively, from high resolution UVES
spectra, Carretta et al. 2009c, and this work), and the HB morphology
significantly
differs among these clusters, as evident from the CMDs from the ACS survey
(Sarajedini et al. 2007) plotted in the lower panels of Fig.~\ref{f:naohb}.
The HB morphological index (defined as (B-R)/(B+V+R) from the number of stars
B,V,R bluer than, inside, and redder than the instability strip, Lee 1990) is
0.93, 0.98 and -0.06 for NGC~288, M~80, and M~4, respectively (Mackey and van
den Bergh 2005), corresponding to the differences clearly visible in the CMDs of
the lower panels in Fig.~\ref{f:naohb}. The HB stellar distribution in the first
two GCs is all to the blue of the RR Lyrae instability strip, starting almost at
the same colour, but reaching far more fainter (hotter) levels in M~80 than in
NGC~288. On the other hand, the blue part of the HB in M~4 barely ends at the
colour where the HB distribution starts in the two other GCs.

\begin{figure*}
\centering
\includegraphics[bb=15 15 570 500, clip, scale=0.52]{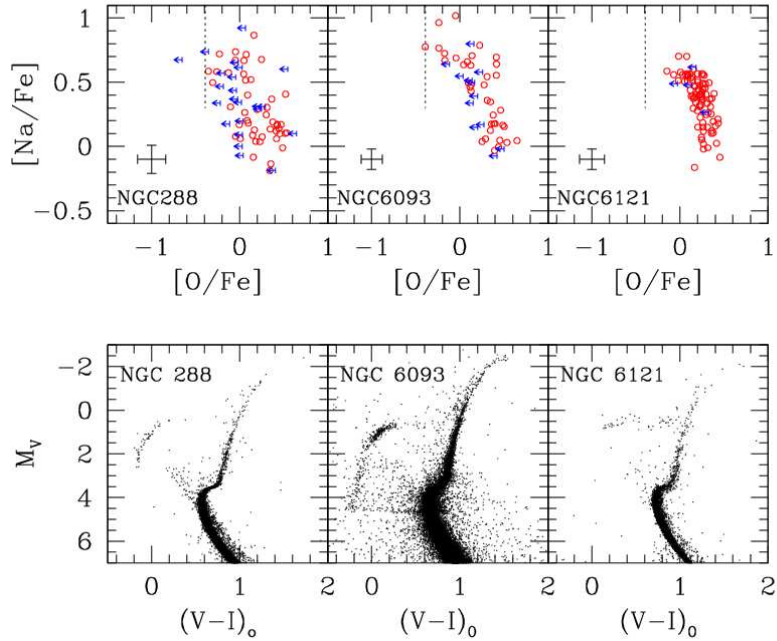}
\caption{Upper panels, from left to right: Na-O anticorrelation for NGC~288
(Carretta et al. 2009a,b), M~80 (this work), and M~4 (NGC~6121, Carretta et al.
2009a,b). The vertical dotted line indicates the reach of the lowest [O/Fe] ratio
measured in M~80, and it is superimposed to the two other GCs for comparison.
Lower panels: $M_V$ vs $V-I$ CMDs for the three GCs from the ACS survey by
Sarajedini et al. (2007). Distance moduli and reddening values are from Harris
(1996).}
\label{f:naohb}
\end{figure*}

\begin{figure*}
\centering
\includegraphics[bb=19 146 587 500, clip, scale=0.52]{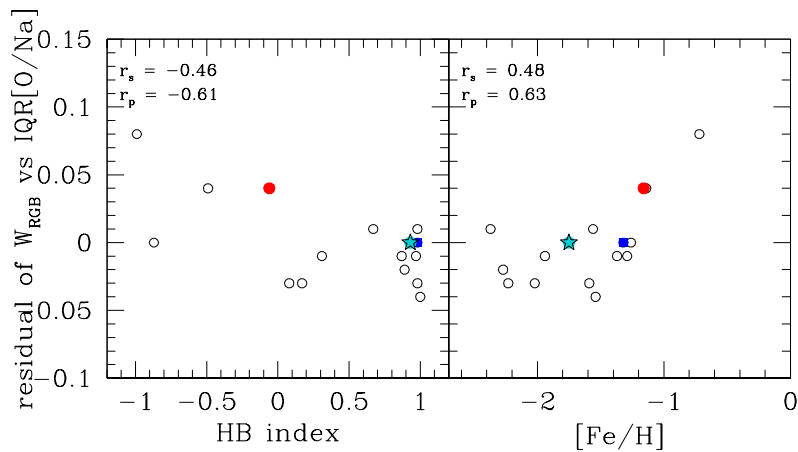}
\caption{Residuals of the $W_{RGB}$ vs IQR[O/Na] relation as a function of the
HB index and [Fe/H] ratio for the 16 GCs in common with the SUMO survey. Symbols
are as in Fig.~\ref{f:sumo}.}
\label{f:sumores}
\end{figure*}

To better understand the missing piece of information from Fig.~\ref{f:sumo} we
then computed the residuals about the relation between $W_{RGB}$ and IQR[O/Na]. 
These value are shown in Fig.~\ref{f:sumores} as a function of the HB index and
metallicity, and allow to better understand the difference between photometric and
spectroscopic properties of multiple populations. The residuals are 
correlated with HB index and [Fe/H] (which is the $first$ parameter controlling
the star distribution on the HB): both relations are found statistically
significant at a level of confidence of about 99\%. 
We conclude that star-to-star abundance variations in the lightest 
proton-capture elements, like C and N, are easily revealed as enlargements in
the photometric sequences in the CMDs, through the impact of variations in the
abundance of N, in particular, on the absorption molecular bands in the
ultraviolet or most bluer filters. These variations also include contributions 
not due to multiple populations but to the standard evolution of population II
low mass stars: the first dredge-up and the second mixing episode after the RGB
bump both concur to enhance the N abundance, equally in first and second
generation stars. From the study of Monelli et al. (2013), at face value the
changes in the abundances of C and N are comparable in extent for GCs like M~80
and M~4 and NGC~288.

\begin{figure}
\centering
\includegraphics[scale=0.40]{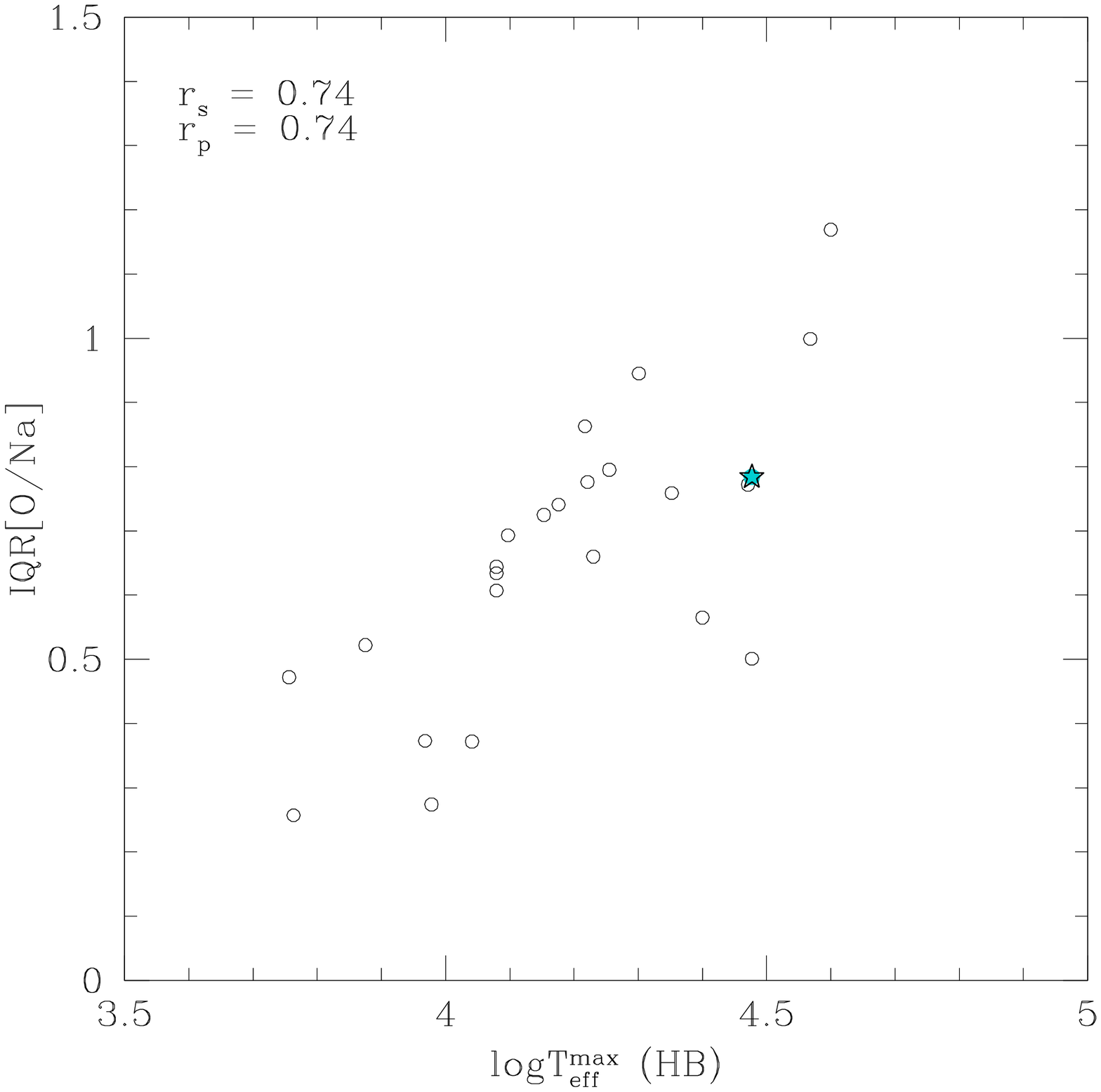}
\caption{The extension of the Na-O anti-correlation (measured using
IQR[O/Na]) as a function of the maximum temperature along the HB, taken from
Recio-Blanco et al. (2006). M~80 is represented by a filled star symbol. The
Spearman rank and the Pearson linear correlation coefficients are listed.}
\label{f:iqrteff60}
\end{figure}

Species like O and Na cannot heavily affect the broad band filters, and this
explain the somewhat loose correlation in Fig.~\ref{f:sumo}. Second generation
stars with O-poor/Na-rich composition are also expected to show N enhancement
(including the contribution from mere evolutive mixing), and this gives the
correlation with the width of the RGB in index like $c_{U,B,I}$. On the other
hand, when O and Na are more changed with respect to the primordial composition
of stars, also He is more and more modified. Taking into account this occurrence
results into a tighter correlation between photometric and spectroscopic
index, and produces the significant correlation with HB morphology, for which He
is the third parameter (see Gratton et al. 2010 and references therein).
As a sanity check, if O and Na are more related to He and the HB morphology we
should expect that M~80 participates to the tight correlation between the
extension of the Na-O anticorrelation and the bluest and hottest point reached
along the HB, discovered by Carretta et al. (2007b). This is just the case, as
shown in Fig.~\ref{f:iqrteff60}, where we used for M~80 the value 
$\log T_{\rm eff}^{max}=4.477$ from Recio-Blanco et al. (2006). M~80 nicely
falls on the relation defined by the strict correspondence between the
phenomenon of multiple stellar populations in GCs and their HB morphology.

\subsection{Other elements}

Elements produced with $\alpha-$capture in explosive nucleosynthesis of
core-collapse supernovae (Si, Ca, Ti) track each other in M~80. The observed
scatter in Mg is from 2 to 3 times larger than that of the other
$\alpha-$elements because Mg is also involved in the proton-capture reactions
which change the chemical composition of second generation stars with respect to
the level established by Type II SNe. 

\begin{figure}
\centering
\includegraphics[scale=0.45]{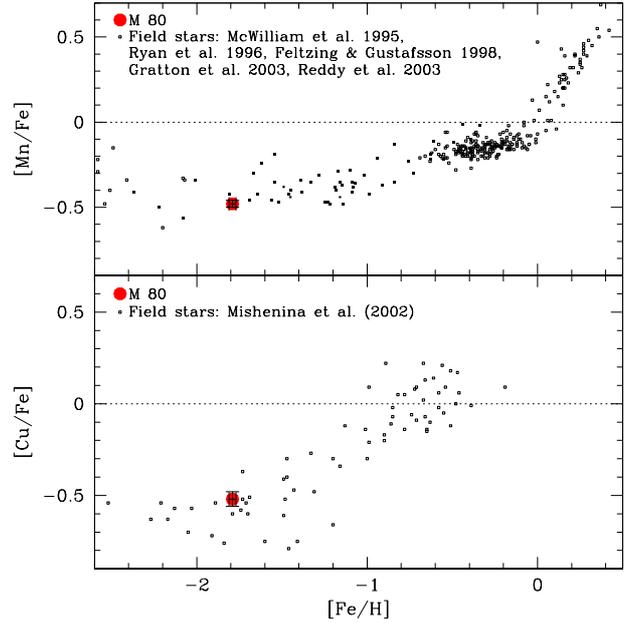}
\caption{[Mn/Fe] (upper panel) and [Cu/Fe] (lower panel) average ratios of M~80
from stars with UVES spectra compared to various samples of field stars as a
function of metallicity. The $rms$ scatter for Mn, Cu and Fe are plotted for
M~80 (they are inside the large symbol).}
\label{f:mncu60}
\end{figure}

The pattern of Fe-group elements in M~80 is not unusual; these elements nicely
track Fe, apart from Mn and Cu, which are found to be  underabundant in
excellent agreement with the measurements in field  stars of the Galactic halo
(see Fig.~\ref{f:mncu60}).

We conclude that concerning the abundance pattern of species from proton- 
and $\alpha-$capture, as well as those of the Fe-peak, M~80 is a typical halo 
cluster.

\begin{figure}
\centering
\includegraphics[bb=19 146 418 709, clip, scale=0.52]{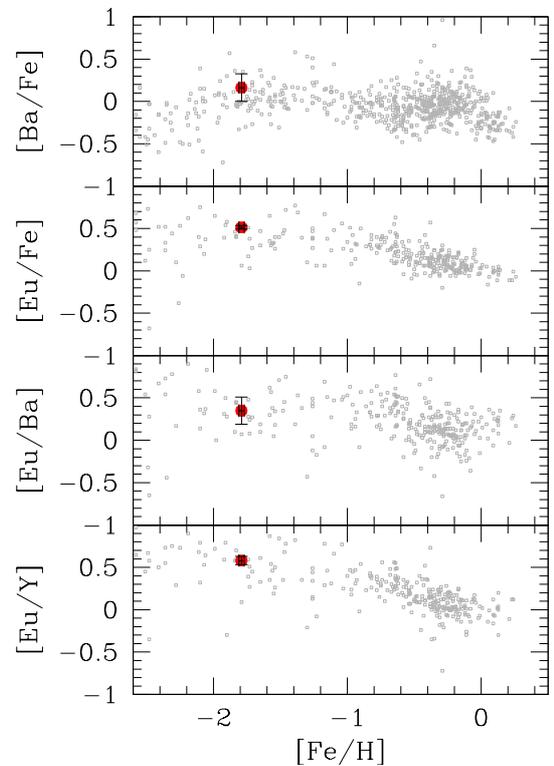}
\caption{From top to bottom: [Ba/Fe], [Eu/Fe], [Eu/Ba], and [Eu/Y] abundance
ratios in M~80 (large filled circle) and in Galactic field stars (Venn et al.
2004, empty squares), as a function of metallicity.}
\label{f:baeuy60}
\end{figure}

Also the pattern of the abundances of neutron-capture elements looks
normal in M~80. In Fig.~\ref{f:baeuy60} we show the average abundance ratios of
a light neutron-capture element, Y, of the
heavier element Ba, created by the main $s-$process, and of the Eu forged almost
totally in the $r-$process (Arlandini et al. 1999). The mean ratios in M~80 are
found compatible to the run of these elements in field stars of the
Galactic halo as a function of metallicity (Venn et al. 2004). We do not confirm
the very high value [Eu/Fe]$=+0.80$ dex found by Cavallo et al. (2004), which
made M~80 outstanding in the plot showed by Venn et al.

To characterize the modality of enrichment in heavy elements from neutron
capture it is safer to use the La abundance rather than Ba: although derived for
the limited sample of stars with UVES spectra, La lines are weak and less
sensitive to uncertainty in the abundance analysis. The comparison between La (a
species with predominance of the $s-$process in the solar system) and Eu (almost
totally $r-$process element in solar system) allows to study the aforementioned 
enrichment. We found in M~80 a mean value of 
$\log \epsilon$(La/Eu)=0.44 dex ($\sigma=0.11$ dex, 14 stars) which would
indicate a predominance from $r-$process production in this cluster (pure $r$
level is 0.09 dex, Simmerer et al. 2004).

\begin{figure}
\centering
\includegraphics[bb=30 175 520 704, clip, scale=0.52]{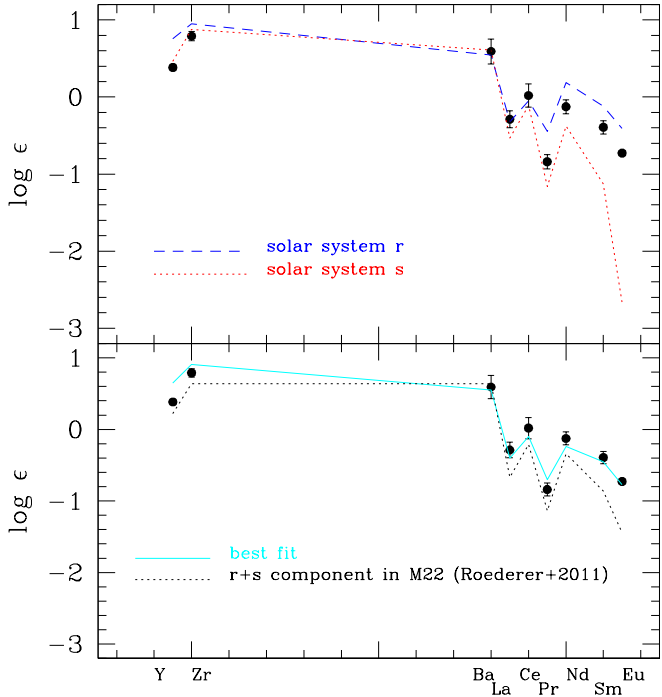}
\caption{Average abundances of neutron-capture elements Y, Zr, Ba, La, Ce, Pr, 
Nd, Sm, and Eu in M~80. The associated $rms$ scatters are also shown. In the
upper panel the solar scaled abundance of pure $r-$process (dashed line) and of
pure $s-$process (dotted line) from Simmerer et al. (2004) are shown. In the
lower panel the attempt to best fit the observed abundances with solar scaled
pure $r-$process and $s-$process abundances (Simmerer et al. 2004) is shown
(light blue line): the scaling factor adopted are -1.27 dex and -1.81 dex for
the $r-$  and $s-$process, respectively. The black dotted line connects the
abundances in the $r+s$ group of the GC M~22 derived by Roederer et al. (2011).
}
\label{f:ncapt60}
\end{figure}

A more detailed comparison can be performed by comparing all the available
ratios of neutron-capture elements in M~80 with the pattern of solar scaled 
contributions from $s-$ and $r-$process nucleosynthesis. In the upper panel of
Fig.~\ref{f:ncapt60} the comparison is made using the fractions estimated in
Simmerer et al. (2004). Several elements occupy a position intermediate between
the pure $s-$process level and the pure $r-$ fraction (obtained as usual from 
$N_{SS,r} \equiv N_{SS,total} - N_{SS,s}$).

Therefore, we tried to obtain a best fit by trying to reproduce the composition 
of Tab.~\ref{t:meanabu60} as the sum of two contributions (a pure solar-scaled
$r-$process and a pure solar scaled $s-$process: $s-$ and $r-$ fractions are
as in Simmerer et al. 2004) with suitable scaling factors. We found that the 
observed pattern can be reproduced by
using scaling factors of -1.27 dex and -1.81 dex for the $r-$ and $s-$process,
respectively (see lower panel of Fig.~\ref{f:ncapt60}). Hence, if we consider 
that [Fe/H]$=-1.79$ dex, this implies
[$r$/Fe]=+0.52 and [$s$/Fe]=-0.02 dex. The excess of elements produced by
the $r-$process is similar to that obtained for the $\alpha-$elements and
can be interpreted as an Fe deficiency due to the fact that there
was no significant contribution of SN Ia to the material from which M~80
formed. On the other hand, a significant contribution by the $s-$process
is required to explain observations. The abundance ratio first/second 
peak of the $s-$process is more favorable to the second. This is not
unexpected, given the low metallicity of M80.

To get more insight with a direct, model independent approach we 
compared the pattern found in M~80 with that of two other GCs, one showing a
dispersion in the $r-$process content and another one where a clear dispersion
in $s-$process elements has been detected, chosen from the extensive survey by 
Roederer (2011).

\begin{figure}
\centering
\includegraphics[scale=0.40]{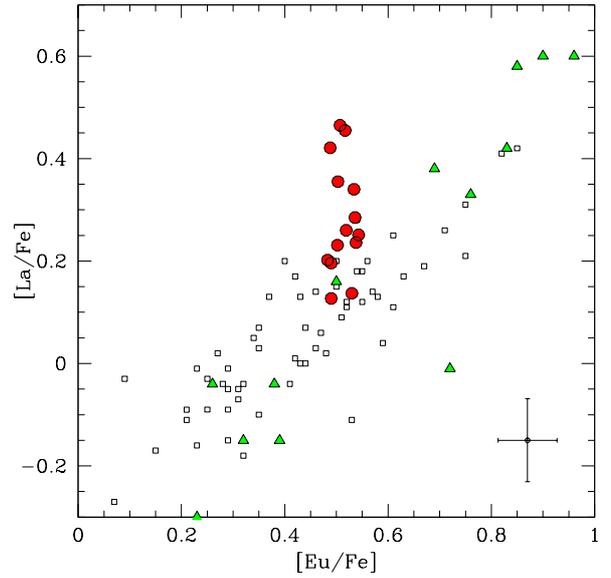}
\caption{[La/Fe] ratios as a function of [Eu/Fe] ratios for individual stars in
M~80 (large filled circles) compared to $r-$only field stars from Roederer et
al. (2010; small empty squares) and to a GC with a dispersion in $r-$process
from Roederer (2011): M~15 (filled triangles), with abundances from Otsuki et
al. (2006) and Sobeck et al. (2011). The internal error bars refer to our
analysis of M~80.}
\label{f:eula60r}
\end{figure}

In Fig.~\ref{f:eula60r} our ratios [Eu/Fe] and [La/Fe] for individual stars in
M~80 are compared to the sample of field stars with no detectable $s-$process
contribution analyzed by Roederer et al. (2010). Superimposed to these we also
plot stars in M~15, a GC well know to show a dispersion in the $r-$process
content (see Roederer 2011). For M~15 we used data from Otsuki et al. (2006)
and Sobeck et al. (2011).

The M~15 data show the same correlation between [La/Fe] and [Eu/Fe] present in
the sample of {\it bona fide} $r-$only metal-poor field stars from the
compilation of Roederer et al. (2010). A dispersion in [La/Fe] was found to be
intrinsic to this sample, indicating that these variations are intrinsically
associated to the $r-$process yields (Roederer et al. 2010). On the other hand,
it is immediately evident that no intrinsic dispersion in Eu is present among
giants in M~80, whereas we cannot exclude a possible small 
spread of La abundances in these stars.

\begin{figure}
\centering
\includegraphics[scale=0.40]{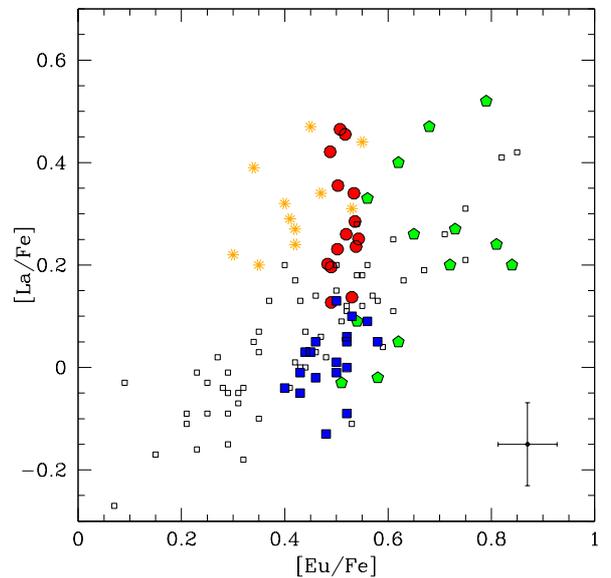}
\caption{[La/Fe] ratios as a function of [Eu/Fe] ratios for individual stars in
M~80 (large filled circles) compared to $r-$only field stars from Roederer et
al. (2010; small empty squares) and to GCs with a dispersion in $s-$process
elements. Blue filled squares indicate the $r-$only component in M~22 from
Marino et al. (2011), orange asterisks are used for the $r+s$ component in M~22
(see Roederer et al. 2011, Marino et al. 2011), and green filled pentagons are
giants in NGC~1851 from Carretta et al. (2011b). The internal error bars refer 
to our present analysis of M~80.}
\label{f:eula6018}
\end{figure}

In Fig.~\ref{f:eula6018} we compare abundances of La
and Eu in M~80 with those of GCs known to host stellar populations with an
intrinsic dispersion in $s-$process elements: M~22 (Marino et al. 2011, Roederer
et al. 2011) and NGC~1851 (Carretta et al. 2011b). The $r-$only component in
M~22 (in the terminology by Roederer et al. 2011) nicely falls on the locus of
galactic field metal-poor stars. This behaviour is shared also by the majority
of stars analyzed in NGC~1851 and by a small fraction of the present sample in 
M~80.
However, most giants in M~80 have distinctly higher La abundances,
and populate a region in the [La/Fe] vs [Eu/Fe] plane where are located stars of
the so-called $r+s$ component in M~22 (Roderer et al. 2011; they correspond to
the $s-$rich group in Marino et al. 2011). This trend extends to higher
metallicities thanks to a minority component in NGC~1851.

The spread observed in La for RGB stars in M~80 seems comparable to the range
observed in the $r+s$ group in M~22. On the other hand, although similarities 
between M~22 and M~80 include the total mass and metal abundance 
([Fe/H]$\sim -1.79$ dex in both), M~80 does not show a spread in [Fe/H] as
instead observed in M~22 (Da Costa et al. 2009, Marino et al. 2011).

Anyway, if we plot the contribution from the $r+s$ group in M~22 from the
analysis by Roederer et al. (2011) over the distribution of neutron-capture
elements observed in M~80 (dotted line in the lower panel of
Fig.~\ref{f:ncapt60}, the agreement is not dramatically unsatisfactory.

We conclude that M~80 could be a less extreme case of a GC with a
significant contribution from the $s-$process, and may be more similar to
NGC~1851, although at lower metallicity.
Although most of neutron-capture elements are only available for the limited
sample of stars with UVES spectra, we found no statistically significant 
difference in the level of abundances of heavy elements between different
stellar generations in M~80.

\section{Summary}

NGC~6093=M~80 is the last of the 24 globular clusters of our core program on the
Na-O anticorrelation in RGB stars and its relation with HB morphology and
cluster parameters (see e.g., Carretta et al. 2006, 2009a,b). In this paper we
present the analysis of this moderately massive, metal-poor, very dense GC, seen
in the direction of the Galactic centre. We used FLAMES spectra of 82 red
giant stars, confirmed members on the basis of their RV and metallicity and 
derived atmospheric parameters, metallicity, and chemical
abundances (12 species for GIRAFFE spectra, more than 20 for UVES spectra).

The main goal is the determination of O and Na abundances, to study the shape
and extension of the Na-O anticorrelation, the main chemical signature of
multiple populations in GCs (e.g., Carretta et al. 2010a). The ratios between
first and second generation stars is normal, with fractions of P, I, and E 
stars at 36, 56, and 8\%, respectively. The three groups
of stars appear to isolate naturally  (see Fig. 8). The spread of both O and Na
is large, about 1 dex, and the anticorrelation has a moderate extension (as
measured by a IQR[O/Na]=0.784), which places NGC~6093 in the middle of the
correlation with cluster mass (via $M_V$, see Fig. 11) and maximum temperature
along the HB (see Fig. 16). The cluster also conforms to the relation between
concentration and the residual around the best fit IQR[O/Na] and $M_V$.   

Using the UVES spectra of 14 member stars we can also study
other p-capture elements. We find that Al is enhanced in I, E stars and is
anticorrelated with Mg, as usual. Briefly, this inner-halo cluster shows the
classical chemical signatures of multiple stellar populations in the normal
ways.

All $\alpha$, proton-capture, and iron peak elements behave in NGC~6093 like in
a typical halo cluster. The neutron-capture elements (with the exception of Ba)
can be measured only in the UVES spectra. They conform to the run of field stars
as function of metallicity (see Fig. 18). Comparing the values of [La/Fe] and
[Eu/Fe] with those of two clusters showing some spread in $r-$process or $s-$process
elements (M 15 and M 22, respectively) we see that NGC~6093 may be a mild case
of intrinsic dispersion is $s-$process elements (see Fig. 21), without showing,
however, any indication of metallicity spread.

\begin{acknowledgements}
This publication makes use of data products from the Two Micron All Sky Survey,
which is a joint project of the University of Massachusetts and the Infrared
Processing and Analysis Center/California Institute of Technology, funded by the
National Aeronautics and Space Administration and the National Science
Foundation.  This research has been funded by PRIN INAF 2011
"Multiple populations in globular clusters: their role in the Galaxy assembly"
(PI E. Carretta), and PRIN MIUR 2010-2011, project ``The Chemical and Dynamical
Evolution of the Milky Way and Local Group Galaxies'' (PI F. Matteucci) . This
research has made use of the SIMBAD database, operated at CDS, Strasbourg,
France and of NASA's Astrophysical Data System.
\end{acknowledgements}

\begin{appendix}

\section{Error estimates}

We refer the reader to the analogous Appendices in Carretta et al. (2009a,b) for
a detailed discussion of the procedure adopted for error estimates. 

\begin{table*}
\centering
\caption[]{Sensitivities of abundance ratios to variations in the atmospheric
parameters and to errors in the equivalent widths, and errors in abundances for
stars of NGC~6093 observed with UVES.}
\begin{tabular}{lrrrrrrrr}
\hline
Element     & Average  & T$_{\rm eff}$ & $\log g$ & [A/H]   & $v_t$    & EWs     & Total   & Total      \\
            & n. lines &      (K)      &  (dex)   & (dex)   &kms$^{-1}$& (dex)   &Internal & Systematic \\
\hline        
Variation&             &  50           &   0.20   &  0.10   &  0.10    &         &         &            \\
Internal &             &   4           &   0.04   &  0.02   &  0.10    & 0.01    &         &            \\
Systematic&            &  58           &   0.06   &  0.08   &  0.03    &         &         &            \\
\hline
$[$Fe/H$]${\sc  i}& 56 &    +0.065     & $-$0.005 &$-$0.013 & $-$0.018 & 0.011  &0.022    &0.076	\\
$[$Fe/H$]${\sc ii}&  8 &  $-$0.033     &   +0.085 &  +0.023 & $-$0.007 & 0.029  &0.035    &0.046	\\
$[$O/Fe$]${\sc  i}&  2 &  $-$0.048     &   +0.083 &  +0.041 &	+0.015 & 0.057  &0.062    &0.088	\\
$[$Na/Fe$]${\sc i}&  3 &  $-$0.022     & $-$0.047 &$-$0.020 &	+0.012 & 0.047  &0.049    &0.066	\\
$[$Mg/Fe$]${\sc i}&  2 &  $-$0.020     & $-$0.011 &$-$0.002 &	+0.008 & 0.057  &0.058    &0.035	\\
$[$Al/Fe$]${\sc i}&  2 &  $-$0.019     & $-$0.010 &$-$0.001 &	+0.016 & 0.057  &0.060    &0.194	\\
$[$Si/Fe$]${\sc i}&  7 &  $-$0.056     &   +0.023 &  +0.014 &	+0.014 & 0.031  &0.034    &0.066	\\
$[$Ca/Fe$]${\sc i}& 16 &    +0.008     & $-$0.020 &$-$0.009 & $-$0.009 & 0.020  &0.023    &0.012	\\
$[$Sc/Fe$]${\sc ii}& 8 &    +0.030     & $-$0.016 &  +0.001 & $-$0.008 & 0.029  &0.030    &0.036	\\
$[$Ti/Fe$]${\sc i}&  9 &    +0.029     & $-$0.015 &$-$0.008 &   +0.008 & 0.027  &0.028    &0.036	\\
$[$Ti/Fe$]${\sc ii}& 9 &    +0.024     & $-$0.024 &$-$0.007 & $-$0.011 & 0.027  &0.030    &0.029	\\
$[$V/Fe$]${\sc i} &  8 &    +0.033     & $-$0.014 &$-$0.009 &	+0.004 & 0.029  &0.029    &0.039	\\
$[$Cr/Fe$]${\sc i}& 12 &    +0.048     & $-$0.021 &$-$0.014 & $-$0.023 & 0.023  &0.033    &0.057	\\
$[$Cr/Fe$]${\sc ii}& 7 &    +0.002     & $-$0.019 &$-$0.014 &	+0.001 & 0.031  &0.031    &0.009	\\
$[$Mn/Fe$]${\sc i}&  3 &    +0.005     & $-$0.011 &$-$0.003 &	+0.009 & 0.047  &0.048    &0.009	\\
$[$Co/Fe$]${\sc i}&  3 &  $-$0.010     & $-$0.004 &  +0.003 &	+0.015 & 0.047  &0.049    &0.014	\\
$[$Ni/Fe$]${\sc i}& 17 &  $-$0.004     &   +0.010 &  +0.008 &	+0.005 & 0.020  &0.020    &0.008	\\
$[$Cu/Fe$]${\sc i}&  2 &  $-$0.001     &   +0.001 &  +0.001 & $-$0.001 & 0.057  &0.057    &0.011	\\
$[$Zn/Fe$]${\sc i}&  1 &  $-$0.087     &   +0.039 &  +0.019 & $-$0.005 & 0.081  &0.082    &0.103	\\
$[$Y/Fe$]${\sc ii}& 12 &    +0.036     & $-$0.023 &$-$0.004 & $-$0.019 & 0.023  &0.031    &0.044	\\
$[$Zr/Fe$]${\sc i}&  3 &    +0.069     & $-$0.010 &$-$0.012 &	+0.016 & 0.047  &0.050    &0.082	\\
$[$Zr/Fe$]${\sc ii}& 1 &    +0.030     & $-$0.016 &$-$0.002 &	+0.003 & 0.081  &0.081    &0.039	\\
$[$Ba/Fe$]${\sc ii}& 3 &    +0.052     & $-$0.026 &  +0.004 & $-$0.079 & 0.047  &0.092    &0.078	\\
$[$La/Fe$]${\sc ii}& 1 &    +0.058     &   +0.008 &  +0.002 & $-$0.010 & 0.081  &0.082    &0.073	\\
$[$Ce/Fe$]${\sc ii}& 1 &    +0.042     & $-$0.018 &$-$0.001 & $-$0.001 & 0.081  &0.081    &0.063	\\
$[$Pr/Fe$]${\sc ii}& 4 &    +0.045     & $-$0.016 &  +0.000 &   +0.002 & 0.041  &0.041    &0.058	\\
$[$Nd/Fe$]${\sc ii}& 7 &    +0.049     & $-$0.019 &$-$0.001 & $-$0.014 & 0.031  &0.034    &0.061	\\
$[$Sm/Fe$]${\sc ii}& 1 &    +0.052     & $-$0.021 &$-$0.001 & $-$0.003 & 0.081  &0.081    &0.065	\\
$[$Eu/Fe$]${\sc ii}& 2 &    +0.033     & $-$0.008 &  +0.004 &	+0.002 & 0.047  &0.057    &0.039	\\

\hline
\end{tabular}
\label{t:sensitivityu60}
\end{table*}

\begin{table*}
\centering
\caption[]{Sensitivities of abundance ratios to variations in the atmospheric
parameters and to errors in the equivalent widths, and errors in abundances for
stars of NGC~6093 observed with GIRAFFE.}
\begin{tabular}{lrrrrrrrr}
\hline
Element     & Average  & T$_{\rm eff}$ & $\log g$ & [A/H]   & $v_t$    & EWs     & Total   & Total      \\
            & n. lines &      (K)      &  (dex)   & (dex)   &kms$^{-1}$& (dex)   &Internal & Systematic \\
\hline        
Variation&             &  50           &   0.20   &  0.10   &  0.10    &         &         &            \\
Internal &             &   4           &   0.04   &  0.02   &  0.36    & 0.02    &         &            \\
Systematic&            &  58           &   0.06   &  0.07   &  0.04    &         &         &            \\
\hline
$[$Fe/H$]${\sc  i}& 19 &    +0.060     & $-$0.009 &$-$0.011 & $-$0.017 & 0.022  &0.065    &0.070	\\
$[$Fe/H$]${\sc ii}&  2 &  $-$0.019     &   +0.079 &  +0.013 & $-$0.005 & 0.067  &0.071    &0.033	\\
$[$O/Fe$]${\sc  i}&  1 &  $-$0.043     &   +0.085 &  +0.032 &	+0.020 & 0.095  &0.121    &0.063	\\
$[$Na/Fe$]${\sc i}&  2 &  $-$0.026     & $-$0.028 &$-$0.006 &	+0.013 & 0.067  &0.082    &0.050	\\
$[$Mg/Fe$]${\sc i}&  1 &  $-$0.025     & $-$0.001 &  +0.001 &	+0.012 & 0.095  &0.104    &0.030	\\
$[$Si/Fe$]${\sc i}&  3 &  $-$0.044     &   +0.021 &  +0.008 &	+0.015 & 0.055  &0.077    &0.052	\\
$[$Ca/Fe$]${\sc i}&  4 &  $-$0.010     & $-$0.009 &$-$0.002 & $-$0.002 & 0.048  &0.048    &0.012	\\
$[$Sc/Fe$]${\sc ii}& 4 &  $-$0.052     &   +0.082 &  +0.027 &   +0.009 & 0.048  &0.060    &0.065	\\
$[$Ti/Fe$]${\sc i}&  3 &    +0.009     & $-$0.004 &$-$0.002 &	+0.013 & 0.055  &0.072    &0.012	\\
$[$V/Fe$]${\sc i} &  4 &    +0.026     & $-$0.007 &$-$0.003 &	+0.016 & 0.048  &0.075    &0.031	\\
$[$Cr/Fe$]${\sc i}&  2 &    +0.006     & $-$0.006 &$-$0.001 &   +0.019 & 0.067  &0.096    &0.013	\\
$[$Ni/Fe$]${\sc i}&  4 &  $-$0.001     &   +0.009 &  +0.004 &	+0.012 & 0.048  &0.064    &0.006	\\
$[$Ba/Fe$]${\sc ii}& 1 &  $-$0.038     &   +0.076 &  +0.026 & $-$0.060 & 0.095  &0.237    &0.067	\\

\hline
\end{tabular}
\label{t:sensitivitym60}
\end{table*}

\end{appendix}

\clearpage

%\Online
\Online
\begin{table*}
\centering \tiny
\setcounter{table}{1}
\caption{List and relevant information for target stars in NGC~6093. The
complete table is available electronically only at CDS.}.
\begin{tabular}{rcccclrrr}
\hline\hline
ID & RA &Dec &$B$ &$V$ &setup &RV$_H$ &err(RV) &dist \\
   & HH:MM:SS &DD:PP:SS & & & & km~s$^{-1}$ & km~s$^{-1}$ &arcmin \\
\hline
\multicolumn{9}{c}{Members on the basis of radial velocity and metallicity} \\
  2174  &16 16 58.972 &-22 53 15.440  &17.113 &16.040 &HR11,13 & 16.45  &0.18& 5.37	\\ 
  2232  &16 17 04.518 &-22 52 50.850  &16.877 &15.760 &HR11,13 & 14.96  &0.03& 5.74	\\ 
 10043  &16 16 42.766 &-23 02 09.290  &16.643 &15.475 &HR13 &  9.74  &0.60& 5.77	\\ 
 11607  &16 17 19.568 &-23 05 43.100  &16.914 &15.771 &HR11,13 & 13.26  &0.47& 8.17	\\ 
 11776  &16 16 52.809 &-23 04 41.900  &15.606 &14.325 &HR11,13 & 14.39  &0.13& 6.52	\\ 
\hline
\end{tabular}
\label{t:coo60}
\end{table*}

\setcounter{table}{2}
\begin{table*}
\centering
\caption[]{Adopted atmospheric parameters and derived iron abundances in NGC~6093.
n is the number of lines used in the analysis. The
complete table is available electronically only at CDS.}
\begin{tabular}{rccccrcccrccc}
\hline
Star   &  $T_{\rm eff}$ & $\log$ $g$ & [A/H]  &$v_t$	     & n  & [Fe/H]{\sc i} & $rms$ & n  & [Fe/H{\sc ii} & $rms$ \\
       &     (K)	&  (dex)     & (dex)  &(km s$^{-1}$) &    & (dex)	  &	  &    & (dex)         &       \\
\hline
 2174	 & 4973& 2.30 & $-$1.73 & 1.70 &  21 & $-$1.729 & 0.118 &  2  & $-$1.697 &0.073  \\  
 2232	 & 4887& 2.12 & $-$1.76 & 1.84 &  22 & $-$1.757 & 0.068 &  1  & $-$1.773 &       \\  
10043	 & 4829& 2.00 & $-$1.78 & 2.11 &  17 & $-$1.776 & 0.077 &  3  & $-$1.787 &0.059  \\  
11607	 & 4905& 2.17 & $-$1.77 & 1.82 &  22 & $-$1.773 & 0.085 &  1  & $-$1.795 &       \\  
11776	 & 4546& 1.46 & $-$1.81 & 0.88 &  37 & $-$1.807 & 0.155 &  2  & $-$1.810 &0.031  \\  

\hline
\end{tabular}
\label{t:atmpar60}
\end{table*}

\setcounter{table}{4}
\begin{table*}
\centering
\caption[]{Abundances of proton-capture elements in stars of NGC~6093.
Upper limits (limO,Al=0)
and detections (=1) for O and Al are flagged. The
complete table is available electronically only at CDS.}  
\begin{tabular}{rrccrccrccrcccc}
\hline
       star  &n &  [O/Fe]&  rms  &  n& [Na/Fe]& rms  &	n& [Mg/Fe]& rms  &  n&[Al/Fe]& rms  &limO&    limAl  \\ 
\hline       
 2174	 &  2 &   +0.14 & 0.13  & 4 &	+0.52 & 0.08  & 3 & +0.51 & 0.10     &   &	  &	 &  0  &    \\     
 2232	 &  1 & $-$0.12 &       & 4 &	+0.64 & 0.04  & 1 & +0.42 &	     &   &	  &	 &  0  &    \\     
10043	 &  2 &   +0.55 & 0.01  &   &	      &       & 1 & +0.44 &	     &   &	  &	 &  1  &    \\     
11607	 &  1 &   +0.15 &       & 2 &	+0.34 & 0.07  & 1 & +0.46 &	     &   &	  &	 &  0  &    \\     
11776	 &  2 &   +0.17 & 0.11  & 3 &	+0.50 & 0.08  & 2 & +0.29 & 0.05     &   &	  &	 &  0  &    \\     
\hline
\end{tabular}
\label{t:proton60}
\end{table*}

\setcounter{table}{5}
\begin{table*}
\centering
\caption[]{Abundances of $\alpha$-elements in stars of NGC~6093. 
The
complete table is available electronically only at CDS.}
\begin{tabular}{rrccrccrccrcc}
\hline
   star      &  n&[Si/Fe]&  rms &    n &  [Ca/Fe]& rms &   n &[Ti/Fe]~{\sc i} &  rms &n &[Ti/Fe]~{\sc ii} & rms \\
\hline   
 2174	&  3  & +0.33 &  0.15 &   7 &	+0.36  & 0.10 &   1  &  +0.15 &       &   &	     &  \\   
 2232	&  4  & +0.39 &  0.16 &   5 &	+0.36  & 0.07 &   4  &  +0.19 &  0.04 &   &	     &  \\   
10043	&  1  & +0.36 &       &   5 &	+0.36  & 0.10 &   4  &  +0.20 &  0.08 &   &	     &  \\   
11607	&  7  & +0.35 &  0.15 &   7 &	+0.36  & 0.08 &   3  &  +0.18 &  0.20 &   &	     &  \\   
11776	&  4  & +0.38 &  0.03 &   5 &	+0.35  & 0.05 &   3  &  +0.18 &  0.18 &   &	     &  \\   
\hline
\end{tabular}
\label{t:alpha60}
\end{table*}

\setcounter{table}{6}
\begin{table*}
\centering
\caption[]{Abundances of Fe-peak elements in stars of NGC~6093. 
The
complete table is available electronically only at CDS.}
\scriptsize
\setlength{\tabcolsep}{1.3mm}
\begin{tabular}{rrccrccrccrccrccrccrccrccrcc}
\hline
      star    & n &[Sc/Fe]~{\sc ii}&rms&n& [V/Fe]  & rms&  n &[Cr/Fe]~{\sc i}&rms&n&n &[Cr/Fe]~{\sc ii}&rms&n& [Mn/Fe] & rms  &   n &[Co/Fe] & rms   &  n  &[Ni/Fe]  & rms  &  n& [Cu/Fe] &rms n& [Zn/Fe] &rms\\
\hline         
 2174	  &  6  &$-$0.03 & 0.08  &    &         &       &    &         &      &    &  & 	&      &    &	      &       &    &	     &       &   3 & $-$0.14 & 0.14 &	  &	    &     &	    &	   \\    
 2232	  &  7  &$-$0.03 & 0.13  &  3 & $-$0.06 & 0.02  &    &         &      &    &  & 	&      &    &	      &       &    &	     &       &   5 & $-$0.15 & 0.20 &	  &	    &     &	    &	   \\    
10043	  &  2  &$-$0.04 & 0.03  &  3 &   +0.00 & 0.17  &    &         &      &    &  & 	&      &    &	      &       &    &	     &       &   3 & $-$0.11 & 0.10 &	  &	    &     &	    &	   \\    
11607	  &  5  &$-$0.02 & 0.06  &  2 & $-$0.06 & 0.21  &    &         &      &    &  & 	&      &    &	      &       &    &	     &       &   4 & $-$0.17 & 0.12 &	  &	    &     &	    &	   \\    
11776	  &  6  &  +0.03 & 0.20  &  4 & $-$0.01 & 0.08  &  3 &   +0.02 & 0.15 &    &  &         &      &    &	      &       &    &	     &       &   6 & $-$0.14 & 0.14 &	  &	    &     &	    &	   \\    
\hline
\end{tabular}
\label{t:fegroup60}
\end{table*}

\setcounter{table}{7}
\begin{table*}
\caption[]{Abundances of $n-$capture elements in stars of NGC~6093 with 
UVES spectra. The
complete table is available electronically only at CDS.}
\scriptsize
\setlength{\tabcolsep}{1.1mm}
\begin{tabular}{rrccrccrccrccrccrccrccrccrcc}
\hline
  star &n  &[Y/Fe]~{\sc ii}&rms  &n& [Zr/Fe]~{\sc i}&rms &n& [Zr/Fe]~{\sc ii}&rms &n&[La/Fe]~{\sc ii}&rms &n&[Ce/Fe]~{\sc ii}&rms&n& [Pr/Fe]~{\sc ii}&rms &n&[Nd/Fe]~{\sc ii}&rms &n& [Sm/Fe]~{\sc ii}& rms&n&[Eu/Fe]~{\sc ii}& rms  \\
\hline    
13474  &13 &$-$0.10	   & 0.12&3&$-$0.07	    &0.07&1&  +0.00	     &    &1&  +0.34	     &    &1&+0.16	     &   &4&  +0.21	     &0.17&9& +0.22          &0.08&1& +0.46	&     &    2& +0.53	     & 0.04 \\ 
13987  &11 &$-$0.04	   & 0.08& &     	    &	 &1&  +0.07	     &    &1&  +0.46	     &    &1&+0.30	     &   &3&  +0.32	     &0.11&7& +0.31          &0.09&1& +0.46	&     &    1& +0.52	     &      \\ 
14148  & 9 &  +0.02	   & 0.10& &    	    &	 &1&  +0.00	     &    &1&  +0.47	     &    &1&+0.41	     &   &4&  +0.31	     &0.15&6& +0.31          &0.09&1& +0.41	&     &    2& +0.51	     & 0.06 \\ 
14656  &13 &$-$0.08	   & 0.13&1&  +0.04	    &	 &1&  +0.09	     &    &1&  +0.42	     &    &1&+0.28	     &   &4&  +0.34	     &0.16&7& +0.35          &0.09&2& +0.39	& 0.06&    2& +0.49	     & 0.04 \\ 
15308  &12 &$-$0.15	   & 0.08&3&$-$0.10	    &0.04&1&$-$0.09	     &    &1&  +0.20	     &    &1&+0.07	     &   &4&  +0.10	     &0.16&6& +0.11          &0.04&2& +0.32	& 0.05&    2& +0.48	     & 0.08 \\ 
\hline
\end{tabular}
\label{t:neutron60}
\end{table*}

\setcounter{table}{8}
\begin{table*}
\centering
\caption[]{Abundances of Ba~{\sc ii} in stars of NGC~6093. 
The
complete table is available electronically only at CDS. }
\begin{tabular}{rrcc}
\hline
   star  &  n &  [Ba/Fe]{\sc ii} & rms  \\
\hline   
 2174	  &  1 &   +0.22&   \\
 2232	  &  1 & $-$0.04&   \\
10043	  &  1 &   +0.03&   \\
11607	  &  1 &   +0.09&   \\
11776	  &  1 &   +0.57&   \\
\hline
\end{tabular}
\label{t:ba60}
\end{table*}

\end{document}